\documentclass[%
preprint,
 amsmath,amssymb,
 aps,
]{revtex4-1}

\usepackage{dsfont}

\usepackage{graphicx}
\usepackage{dcolumn}
\usepackage{bm}
\usepackage{bbm}

\newcommand{\be}{\begin{equation}}
\newcommand{\ee}{\end{equation}}

\newcommand{\ber}{\begin{eqnarray}}
\newcommand{\eer}{\end{eqnarray}}

\newcommand{\us}{\underline{s}}
\newcommand{\ux}{\underline{\sigma}}

\newcommand{\uz}{\underline{z}}
\newcommand{\uxx}{\underline{x}}

\newcommand{\qssr}{Q^{a b}_{ss}}
\newcommand{\qxxr}{Q^{\alpha \beta}_{\sigma \sigma}}
\newcommand{\qssxr}{Q^{1 a \alpha}_{ss \sigma}}

\newcommand{\qeer}{Q^{\alpha \beta}_{\hat{\eta} \hat{\eta} } }
\newcommand{\qmmr}{Q^{\alpha \beta}_{\hat{\mu} \hat{\mu} } }
\newcommand{\qemr}{Q^{\alpha \beta}_{\hat{\eta} \hat{\mu} } }

\newcommand{\chmr}{C^{1 \alpha}_{\hat{h} \hat{\mu} } }
\newcommand{\cher}{C^{1 \alpha}_{\hat{h} \hat{\eta} } }

\newcommand{\qesr}{Q^{\alpha a}_{\hat{\eta}  s}}
\newcommand{\qmsr}{Q^{\alpha a}_{\hat{\mu}  s}}
\newcommand{\qhsr}{Q^{1 a}_{\hat{h} s} }

\newcommand{\qesxr}{Q^{\alpha 1 \beta}_{\hat{\eta}  s \sigma}}
\newcommand{\qmsxr}{Q^{\alpha 1 \beta}_{\hat{\mu}  s \sigma}}
\newcommand{\chsxr}{C^{1 1 \alpha}_{\hat{h} s \sigma} }

\newcommand{\mxr}{m^{\alpha}_{\sigma}}

\newcommand{\hqssr}{\hat{Q}^{a b}_{ss}}
\newcommand{\hqxxr}{\hat{Q}^{\alpha \beta}_{\sigma \sigma}}
\newcommand{\hqssxr}{\hat{Q}^{1 a \alpha}_{ss \sigma}}

\newcommand{\hqeer}{\hat{Q}^{\alpha \beta}_{\hat{\eta} \hat{\eta} } }
\newcommand{\hqmmr}{\hat{Q}^{\alpha \beta}_{\hat{\mu} \hat{\mu} } }
\newcommand{\hqemr}{\hat{Q}^{\alpha \beta}_{\hat{\eta} \hat{\mu} } }
\newcommand{\hchmr}{\hat{C}^{1 \alpha}_{\hat{h} \hat{\mu} } }
\newcommand{\hcher}{\hat{C}^{1 \alpha}_{\hat{h} \hat{\eta} } }

\newcommand{\hqesr}{\hat{Q}^{\alpha a}_{\hat{\eta} s}}
\newcommand{\hqmsr}{\hat{Q}^{\alpha a}_{\hat{\mu} s}}
\newcommand{\hqhsr}{\hat{Q}^{1 a}_{\hat{h}  s}}

\newcommand{\hqesxr}{\hat{Q}^{\alpha 1 \beta}_{\hat{\eta}  s \sigma}}
\newcommand{\hqmsxr}{\hat{Q}^{\alpha 1 \beta}_{\hat{\mu}  s \sigma}}
\newcommand{\hchsxr}{\hat{C}^{1 1 \alpha}_{\hat{h} s \sigma} }

\newcommand{\hmxr}{\hat{m}^{\alpha}_{\sigma}}

\newcommand{\qss}{Q_{ss}}
\newcommand{\qxx}{Q_{\sigma \sigma}}
\newcommand{\qssx}{Q_{ss \sigma}}

\newcommand{\chh}{C_{\hat{h} \hat{h}}}
\newcommand{\qee}{Q_{\hat{\eta} \hat{\eta} } }
\newcommand{\cee}{C_{\hat{\eta} \hat{\eta} } }

\newcommand{\qmm}{Q_{\hat{\mu} \hat{\mu} } }
\newcommand{\cmm}{C_{\hat{\mu} \hat{\mu} } }

\newcommand{\qem}{Q_{\hat{\eta} \hat{\mu} } }
\newcommand{\cem}{C_{\hat{\eta} \hat{\mu} } }

\newcommand{\chm}{C_{\hat{h} \hat{\mu} } }
\newcommand{\che}{C_{\hat{h} \hat{\eta} } }

\newcommand{\qes}{Q_{\hat{\eta}  s}}
\newcommand{\qms}{Q_{\hat{\mu}  s}}
\newcommand{\qhs}{Q_{\hat{h}  s}}

\newcommand{\ces}{C_{\hat{\eta}  s}}
\newcommand{\cms}{C_{\hat{\mu}  s}}
\newcommand{\chs}{C_{\hat{h}  s}}
\newcommand{\chsx}{C_{\hat{h}  s \sigma}}

\newcommand{\qesx}{Q_{\hat{\eta} s \sigma}}
\newcommand{\cesx}{C_{\hat{\eta} s \sigma}}

\newcommand{\qmsx}{Q_{\hat{\mu} s \sigma}}
\newcommand{\cmsx}{C_{\hat{\mu} s \sigma}}

\newcommand{\mx}{m_{\sigma}}

\newcommand{\hqss}{\hat{Q}_{ss}}
\newcommand{\hqxx}{\hat{Q}_{\sigma \sigma}}
\newcommand{\hqssx}{\hat{Q}_{ss \sigma}}

\newcommand{\hchh}{\hat{C}_{\hat{h} \hat{h}}}
\newcommand{\hqee}{\hat{Q}_{\hat{\eta} \hat{\eta} } }
\newcommand{\hcee}{\hat{C}_{\hat{\eta} \hat{\eta} } }

\newcommand{\hqmm}{\hat{Q}_{\hat{\mu} \hat{\mu} } }
\newcommand{\hcmm}{\hat{C}_{\hat{\mu} \hat{\mu} } }

\newcommand{\hqem}{\hat{Q}_{\hat{\eta} \hat{\mu} } }
\newcommand{\hcem}{\hat{C}_{\hat{\eta} \hat{\mu} } }

\newcommand{\hchm}{\hat{C}_{\hat{h} \hat{\mu} } }
\newcommand{\hche}{\hat{C}_{\hat{h} \hat{\eta} } }

\newcommand{\hqes}{\hat{Q}_{\hat{\eta}  s}}
\newcommand{\hqms}{\hat{Q}_{\hat{\mu}  s}}
\newcommand{\hqhs}{\hat{Q}_{\hat{h}  s}}

\newcommand{\hces}{\hat{C}_{\hat{\eta}  s}}
\newcommand{\hcms}{\hat{C}_{\hat{\mu}  s}}
\newcommand{\hchs}{\hat{C}_{\hat{h}  s}}
\newcommand{\hchsx}{\hat{C}_{\hat{h}  s \sigma}}

\newcommand{\hqesx}{\hat{Q}_{\hat{\eta} s \sigma}}
\newcommand{\hcesx}{\hat{C}_{\hat{\eta} s \sigma}}

\newcommand{\hqmsx}{\hat{Q}_{\hat{\mu} s \sigma}}
\newcommand{\hcmsx}{\hat{C}_{\hat{\mu} s \sigma}}

\newcommand{\hmx}{\hat{m}_{\sigma}}

\DeclareMathOperator\erfc{erfc}
\DeclareMathOperator\sgn{sgn}

\begin{document}


\title{Self-Sustained Clusters as Drivers of Computational Hardness in $p$-spin Models}

\author{Jacopo Rocchi}
\email{j.rocchi@aston.ac.uk}
 \affiliation{
 Nonlinearity and Complexity Research Group, Aston University, Birmingham B4 7ET, United Kingdom
}
\author{David Saad}%
 \email{d.saad@aston.ac.uk}
 \affiliation{
 Nonlinearity and Complexity Research Group, Aston University, Birmingham B4 7ET, United Kingdom
}
\author{Chi Ho Yeung}
\email{chyeung@eduhk.hk}
\affiliation{Department of Science and Environmental Studies, The Education University of Hong Kong, Hong Kong}



\date{\today}

\begin{abstract}

While macroscopic properties of spin glasses have been thoroughly investigated, their manifestation in the corresponding microscopic configurations is much less understood. Cases where both descriptions have been provided, such as constraint satisfaction problems, are limited to their ground state properties.
To identify the emerging microscopic structures with macroscopic phases at different temperatures, we study the $p$-spin model with $p\!=\!3$. We investigate the properties of self-sustained clusters, defined as variable sets where in-cluster induced fields dominate over the field induced by out-cluster spins, giving rise to stable configurations with respect to fluctuations. We compute the entropy of self-sustained clusters as a function of temperature and their sizes. In-cluster fields properties and the difference between in-cluster and out-cluster fields support the observation of slow-evolving spins in spin models. The findings are corroborated by observations in finite dimensional lattices at low temperatures.

\end{abstract}

\pacs{Valid PACS appear here}
\maketitle

Spin glass models of disordered systems are characterized by a rich structure of the free-energy landscape and slow dynamics at low temperature.
Mean field analyses~\cite{sherrington1975solvable, parisi1979infinite} typically provide a characterization of the state of the system based on a set of macroscopic order parameters and have provided many interesting and counterintuitive insights~\cite{mezard1990spin, Parisi2007}. Symmetry properties of the resulting order parameters lead to distinct classes of systems termed One-step Replica Symmetric Breaking (1-RSB)~\cite{zamponiRev, castellani2005spin} and Full Replica Symmetry Breaking~\cite{binder1986spin, mezard1990spin} models; the symmetries reflect the organization of states in the free-energy landscape and correspond to an increasingly more complex structure.

Phase transitions in spin-glass systems have been extensively studied within the macroscopic system representation. In particular, models with 1-RSB are common in physics, for instance in structural glass forming liquids~\cite{kirkpatrick1987p,bouchaud1998out,leuzzi2007,gotze2008complex}, as well as in a range of hard-computational problems in computer science, such as Constraint Satisfaction Problems (CSP)~\cite{mezard1990spin, Monasson2007, mezard2009information}. They typically undergo a sequence of structural transitions when the temperature is decreased: while at temperatures above the dynamical transition $T>T_d$ the system is dominated by a paramagnetic (liquid) state; at lower temperatures $T<T_d$ an exponential number (in the number of variables) of TAP (Thouless-Andersson-Palmer) states emerge~\cite{thouless1977solution, monasson1995structural}, leading to a transition beyond which ergodicity breaks. This \emph{dynamical glass transition} is characterized by a non-decaying spin-spin correlation function in disagreement with the static equilibrium zero value~\cite{crisanti1992sphericalp,crisanti1993sphericalp, cugliandolo1993analytical, barrat1996dynamics, montanari2003nature}.  As the temperature decreases further the number of such states, whose logarithm is called complexity, decreases. Eventually, the complexity vanishes at $T_K$, termed the Kauzmann transition in the physics of glass forming liquids and signals a true second-order phase-transition.

While the different temperature regimes are well understood in terms of the (free-) energy landscape, it is much more difficult to describe the manifestation of such changes in microscopic configurations.
Interesting cases where this connection is clearer are CSP, whose solutions are organized in disconnected clusters which contain frozen variables in intermediate regimes before the satisfiability transition~\cite{mezard2003two, krzakala2007gibbs, montanari2008clusters, achlioptas2009random, semerjian2008freezing, braunstein2016large}. Frozen variables take the same value in all the solutions of individual clusters.

Since CSP are often studied in the context of hard optimization problems at $T\!=\!0$, the main external parameter considered is the ratio of constraints to variables $\alpha$ rather than temperature.
While the small $\alpha$ regime can be considered as a paramagnetic (liquid) state, where solutions can be easily found and it is easy to move from one solution to another, the situation changes suddenly at the dynamical transition ratio $\alpha_d$, where solutions are found in disconnected clusters whose number is exponential in the number of variables~\cite{krzakala2007gibbs}.
In general, frozen variables appear for higher $\alpha\!>\!\alpha_d$ values~\cite{montanari2008clusters} with the exception of particular cases such as $k-$XORSAT, $k\!>\!2$, where they appear \emph{at} the dynamical transition $\alpha_d$~\cite{semerjian2008freezing, montanari2008clusters}. Nevertheless, this understanding is limited to optimal solutions, i.e. ground states of CSP, where as their manifestation at non-zero temperatures remains unclear.

In this work we investigate the existence of frozen-like variables in finite-temperature systems. More precisely, we look for clusters of spin variables that exhibit slow dynamics; the mere existence of such clusters is not guaranteed a priori. Somewhat similar problems have been studied in the context of spin glasses on random graphs~\cite{barrat1999time, ricci2000glassy} and in finite dimensional lattices~\cite{roma2006signature, roma2010domain} showing that it is possible to interpret non-equilibrium dynamical properties in terms of structural properties of the ground states of these systems.
These works rely on the notion of \emph{rigidity lattice}~\cite{barahona1982morphology} and the corresponding analyses can be usually done in small systems. Our approach, while aiming at similar goals, relies on a different concept and can be used to analyze spin models via mean field methods.
The central objects of our approach are \emph{Self-Sustained Clusters} (SSC), introduced in the study of the SK model~\cite{yeung2013self}. Pictorially, SSC can be considered as stable components of the system that make relaxation prohibitively slow at low temperature.

Our analysis is carried out within the framework of fully connected Ising $p-$spin model~\cite{derrida1980random} with $p\!=\!3$, whose $T\!=\!0$ limit coincides with the $k-$XORSAT problems with $k\!=\!3$.
While this model belongs to the 1-RSB class, it can be studied in a non-trivial phase by using a simple ansatz for the order parameters, which nevertheless exhibits interesting and non-trivial dynamical properties.
We compute the entropy of these clusters as a function of their size, and characterize their properties for different temperatures. Additionally, we study their stability by computing the distribution of the local field in typical SSC and show that SSC found at low temperatures can be considered as clusters of slow-evolving spins.

\emph{Model:}
The Hamiltonian of the fully connected 3-spin model~\cite{derrida1980random, crisanti1992sphericalp, crisanti1993sphericalp, mezard2003two} with Ising variables is given by
\be
H=-\sum_{i<j<k = 1}^N J_{ijk} \: s_i s_j s_k\:,
\label{eq:hamiltonianpspin}
\ee
where $J_{ijk}$ are i.i.d. Gaussian random variables, with mean $0$ and variance $p! / (2 N^{p-1})$, where $N$ is the number of spins. A brief description of the phase transitions of the model is provided in the Supplemental Material (SM) as well as in~\cite{mezard2009information, crisanti2005complexity, rizzo2013replica}.
Be $\us$ an arbitrary spin configuration and let us use $\sigma$ variables to define the cluster membership per spin. Given a configuration $\us$ and cluster $C$, we assign $\sigma\!=\!+1$ for in-cluster spins and $\sigma\!=\!-1$ for out-cluster spins. To define the notion a SSC, we write the local field $h_i$ acting on spin $s_i$ as the sum of three contributions,
\be
h_i = \frac{1}{2} \sum_{j ,k } J_{ijk} s_j s_k = \frac{1}{2} (u_i + v_i + w_i)\:,
\label{eq:eqheff}
\ee
\ber
\mbox{where,~~}u_i=&&\sum_{j \in C, k \in C} J_{ijk} s_j s_k\:,~~~ v_i=\sum_{j \notin C, k \notin C} J_{ijk} s_j s_k  \nonumber \\
\mbox{~~and~~}
w_i/2 =&& \sum_{j \in C, k \notin C} J_{ijk} s_j s_k \nonumber \:.
\eer
These three field contributions correspond to fields induced from within the cluster (\textit{in-in} contribution, $u_i$), from outside the cluster (\textit{out-out} contribution, $v_i$) and by both in- and out-cluster spins (\textit{in-out} contribution, $w_i$).
A SSC is a group of spins such that, for each spin, the \textit{in-in} contribution dominates the field $h_i$ with respect to all other contributions, such that the following condition is satisfied
\ber                  
u_i^2 > (v_i + w_i)^2 + \epsilon \qquad \forall \: i \in C,
\label{eq:SSCcond}
\eer
where $\epsilon$ is an external positive parameter that can be arbitrarily tuned.
Self-sustained clusters are of interest since for $i \in C$, local fluctuations, giving $O(1/N)$ contributions, do not change the relative importance of $u_i^2$ and $(v_i + w_i)^2$ and, thus, do not change the direction of the corresponding local fields. These clusters are therefore more stable compared to random groups of spins and offer a different perspective on the dynamical slowing down observed at low temperatures.

To count the number of SSC of size $rN$ in a given configuration $\us$ we define the entropic function of $r$
\be
\mathcal{S}(r\:| \us,\{J_{ijk}\})=N^{-1} \: \log \sum_{\ux} \mathds{I}_{\ux}(\us) \: \delta \left(r N - \sum_{i=1}^N \frac{1+\sigma_i}{2}\right)\:,
\label{eq:eqSrs}
\ee
at a given quenched disorder $\{J_{ijk}\}$ and configuration $\us$. We introduced the indicator function
\be
\mathds{I}_{\ux}(\us) = \prod_{i=1}^{N} \left \{ \frac{1-\sigma_i}{2} + \frac{1+\sigma_i}{2} \: \theta \left[ u_i^2 - (v_i + w_i)^2 -\epsilon \right] \right \}\:,
\label{eq:eqInd}
\ee
returning one if and only if $\ux$ defines a SSC in the configuration $\us$. The $\theta(x)$ in Eq.~(\ref{eq:eqInd}) is the Heaviside function, returning one for $x\!>\!0$ and zero otherwise: its role is to select only those realizations $\ux$ for which the condition given by Eq.~(\ref{eq:SSCcond}) holds. Finally, the Dirac delta function in Eq. (\ref{eq:eqSrs}) enforces the size of the clusters to be $rN$.

Equation~(\ref{eq:eqSrs}) gives the logarithm of the number of SSC (entropy of the clusters) per spin in a given configuration $\us$. As we are interested in the number of SSC in a typical configuration, and assuming that $\mathcal{S}(r\:| \us,\{J_{ijk}\})$ is self-averaging with respect to $\us$ and the quenched disorder $\{J_{ijk}\}$, we define
\be
\mathcal{S}_{\beta}(r) = \mathds{E}_J \mathds{E}_{\us} \left[ \: \mathcal{S}(r\:| \us,\{J_{ijk}\})\right]\:,
\label{eq:eqS}
\ee
where $\mathds{E}_J $ denotes the average over the quenched disorder and $ \mathds{E}_{\us} $ is the average over the Boltzmann weight $ Z^{-1} e^{-\beta H(\us)}$. This is the central object of our computation, because the number of SSC $\mathcal{N}_{\beta}(r)$ of size $r$ in a typical configuration at temperature $\beta^{-1}$ is given by
\be
\mathcal{N}_{\beta}(r)=\exp \left[ N \: \mathcal{S}_{\beta}(r) \right] \:.
\label{eq:eqNcluster}
\ee
The number of large clusters is expected to grow as $T$ decreases, signalling the slowing down of the dynamics.

To investigate the stability of SSC, we compute the distribution of the local fields acting on the internal spins. If we consider the SSC of size $rN$, the quantity of interest is the local fields acting on the in-cluster spins,
\be
P_r(h=\lambda)=\mathds{E}_J \mathds{E}_{\us} \mathds{E}^r_{\ux} \left[ \frac{1}{rN}\sum_i \frac{1+\sigma_i}{2} \delta(h_i-\lambda) \right]\:,
\label{Prh}
\ee
where $\mathds{E}^r_{\ux}$ is the average over SSC of size $r$,
\be
\mathds{E}^r_{\ux} \left[ O\left( \ux, \us,\{J_{ijk}\}\right) \right] = \frac{\sum_{\ux}   \mathds{I}^r_{\ux}(\us)   O\left(\ux, \us,\{J_{ijk}\}\right) }{ \sum_{\ux}   \mathds{I}^r_{\ux}(\us)  }\:,
\label{Prh1}
\ee
and the auxiliary function $\mathds{I}^r_{\ux}(\us)$ is given by
\be
\mathds{I}^r_{\ux}(\us)  = \mathds{I}_{\ux}(\us)  \delta \left(r N - \sum_{i=1}^N \frac{1+\sigma_i}{2}\right)\:.	
\label{Prh2}
\ee
An SSC where many spins experience a strong field can be regarded as a cluster of slow-evolving spins, because the probability of spin flips decreases as the absolute value of the local field increases.
The field $P_r(h)$ is supported primarily for small fields at high temperatures and large fields at low temperatures. As shown in the SM this distribution can be obtained from Eq.~(\ref{eq:eqS}).
\begin{figure}[ht]
\includegraphics[width=120mm]{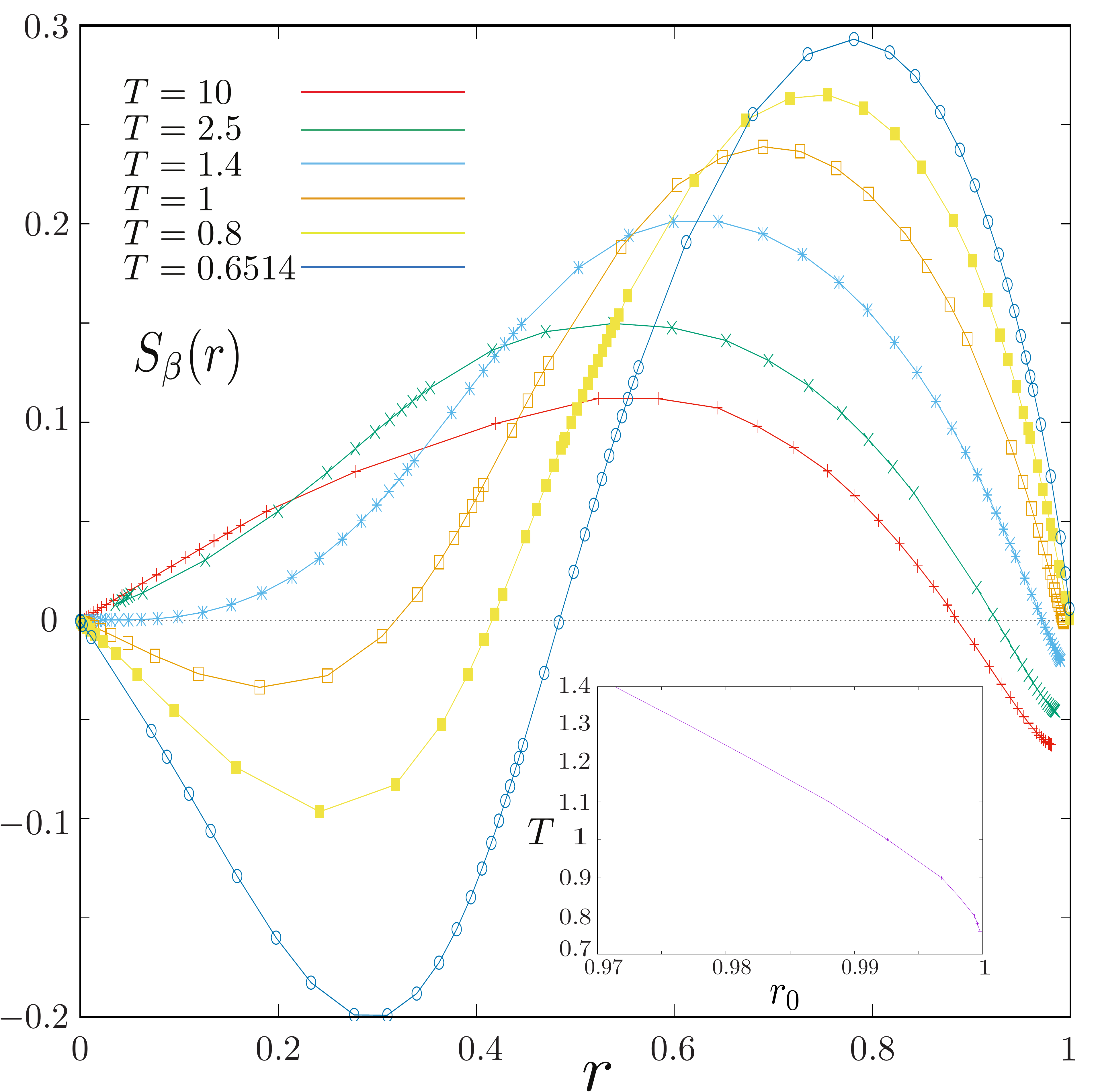}
\caption{\label{fig:entropy} Entropy of the SSC as a function of $r$, where $rN$ is the size of the self sustained cluster, for different temperatures. As the temperature decreases, we observe an increase in the number of large clusters (here, for $\epsilon\!=\!0$). Inset - We observe SSC for cluster size up to $r_0 N$ which increases as $T$ decreases, approaching $1$ for temperatures in the region between $T_d \sim 0.681 $ and $T_{TAP} \sim 0.764$.}
\end{figure}

\emph{Analysis:}
To average over the Boltzmann weights and the disorder and compute $\mathcal{S}_{\beta}(r)$ we will invoke the replica trick twice, once to replace the logarithm in Eq.~(\ref{eq:eqSrs}) and once to account for the partition function in Eq.~(\ref{eq:eqS}). The mathematical identities that we are going to use are:
\be
\log\sum_{\ux} \mathds{I}^r_{\sigma}(\us) =\lim_{m\rightarrow 0} \partial_m \sum_{\ux^{(1)}\ldots \ux^{(m)} } \prod_{\alpha=1}^m \mathds{I}^r_{\ux^{(\alpha)}}\left(\us^{(1)}\right) \:,
\label{eq:eqrep1}
\ee
where $\partial_m$ is the derivative with respect to $m$, $\us^{(1)}\!=\!\us$ and
\be
Z^{-1}=\lim_{n\rightarrow0} \sum_{\us^{(2)}\ldots \us^{(n)}} \exp  \left [ -\beta \sum_{a=1}^n H\left(\us^{(a)}\right) \right ] \:.
\label{eq:eqrep2}
\ee
Expressions are calculated for integers $n$ and $m$ values and then analytically continued to zero~\cite{mezard1990spin}. Greek and Latin indices denote replicas of the $\sigma$ and $s$ variables, respectively.
The details of the computation are discussed in the SM. In this work we employed a Replica Symmetric (RS) ansatz.
In principle, averaging over configurations $\us$, one should invoke a more complex hierarchical ansatz~\cite{mezard1990spin} but the RS ansatz is valid for all temperatures higher than $T_K$, even in the dynamical region between $T_K \sim 0.652$ and $T_d \sim 0.681$ for reasons that can be traced back to the work of Franz and Parisi~\cite{FranzParisi}. In fact, in this regime, the paramagnetic state is replaced by an exponential number of metastable states whose overlap is zero~\cite{kirkpatrick1987dynamics, kurchan1993barriers, crisanti1995thouless}, from which a trivial Parisi function $P(q)=\delta(q)$ is obtained.
We also employ an RS ansatz for the $\sigma$-related order parameters; since $\sigma$ variables are just labels used to define clusters, there is no obvious reason why a more complicated scheme should be invoked.

\begin{figure}[ht]
\includegraphics[width=120mm]{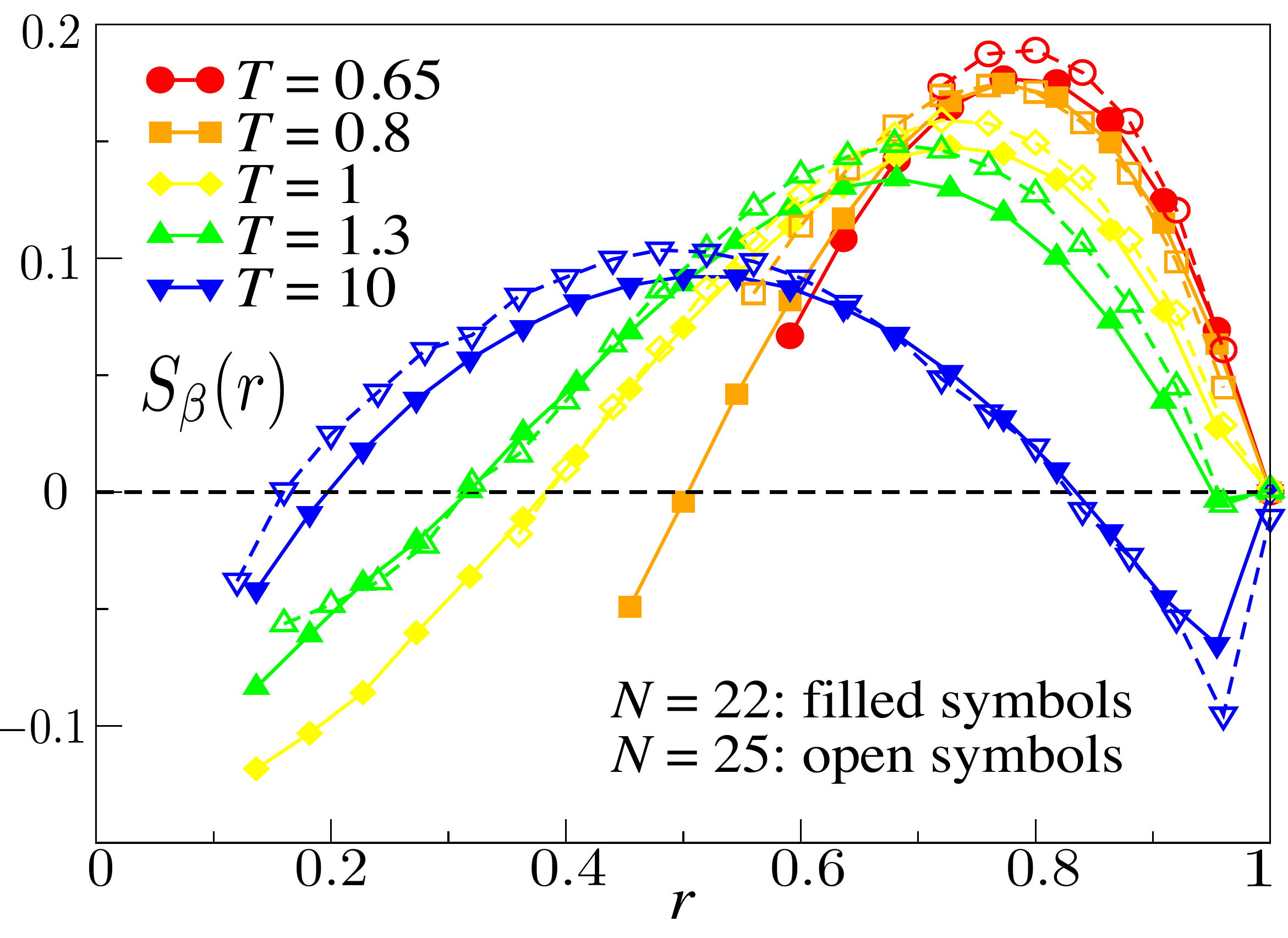}
\caption{\label{fig:small} The entropy of self-sustained clusters in small 3-spin systems. The systems are first initialized with random coupling $J_{ijk}$ and are equilibrated for $5\!\times\! 10^4$ Monte Carlo steps at temperature $T$ before sampling takes place for $1\times 10^3$ steps. The number of SSC is then computed by exhaustive search over all $\underline{\sigma}$ in systems with $N\!=\!22$, and by a random sample of $4\!\times\! 10^6$ of all $\underline{\sigma}$ configurations in systems with $N\!=\!25$; each sampled spin configuration is weighed by the number of times it is sampled. The results are then averaged over 5 realizations of coupling disorders.}
\end{figure}

\emph{Results:} We computed the entropy of SSC for different temperatures and values of $\epsilon$. While we do not observe abrupt changes in the entropy of the SSC, the number of large SSC increases and non-zero entropies appear for larger clusters when $T$ decreases, as shown in Fig.~\ref{fig:entropy}. A numerical analysis performed on small fully connected systems by sampling configurations from Montecarlo, confirm this description, as can be seen in Fig.~\ref{fig:small}.
\begin{figure}[ht]
\includegraphics[width=120mm]{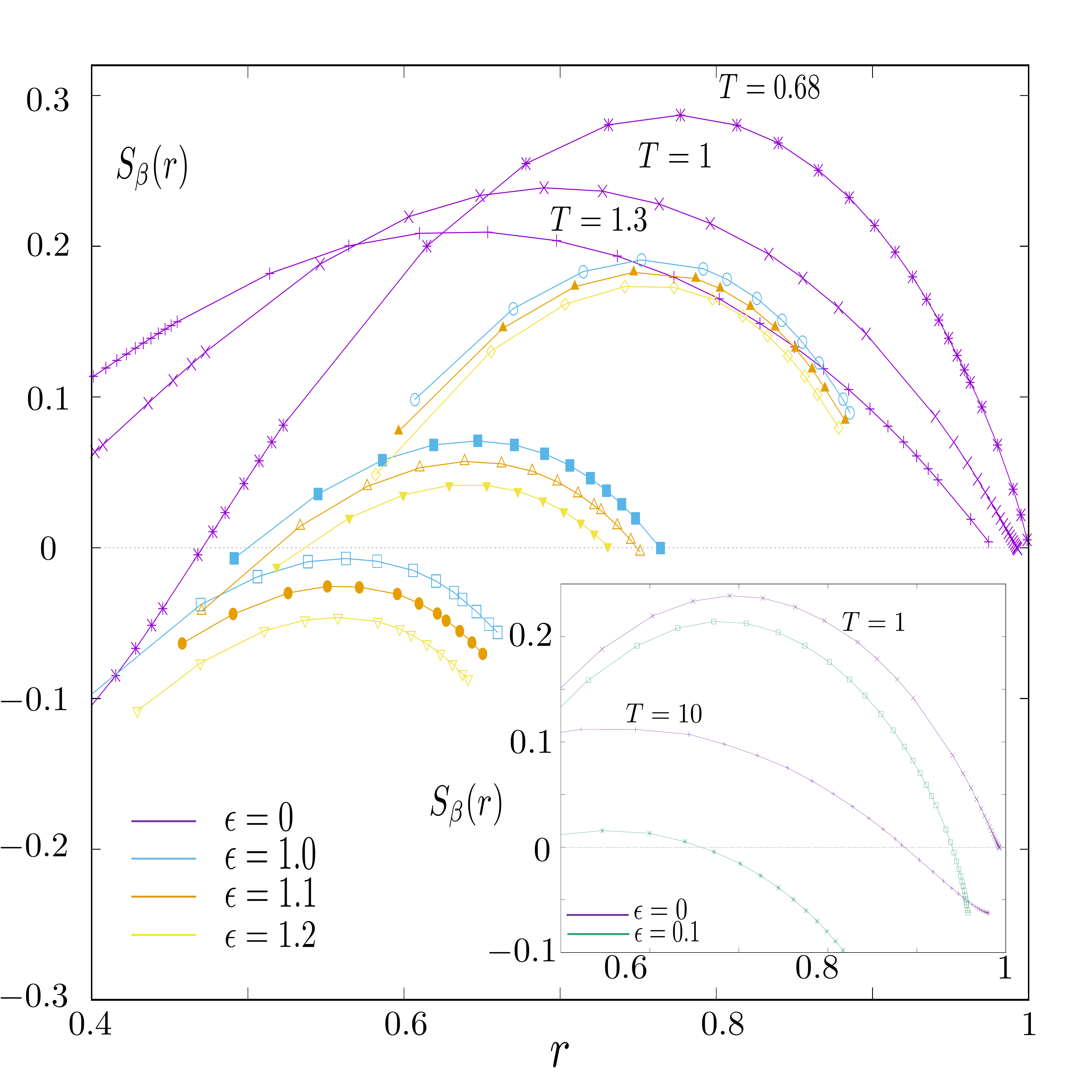}
\caption{\label{fig:eps01} Entropy of the SSC as a function of $r$, where $rN$ is the size of the self sustained cluster, for different temperatures and different $\epsilon$ values. The inset shows the effect of taking the same small $\epsilon$ at two different temperatures: it can be clearly seen that at $T\!=\!10$ it decimates the number of clusters but has relatively little effect at $T\!=\!1$.}
\end{figure}
\begin{figure}[ht]
\includegraphics[width=120mm]{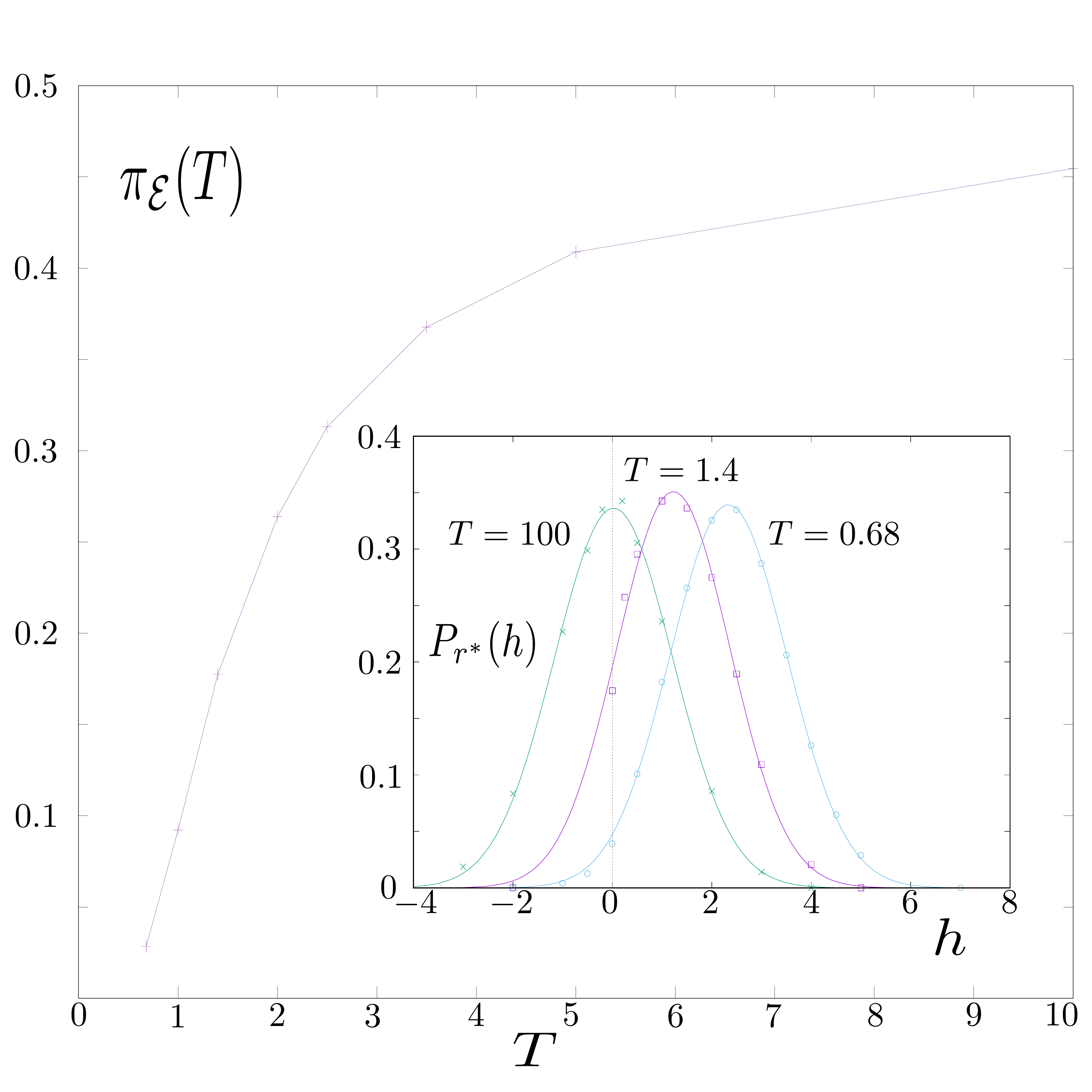}
\caption{\label{fig:Escape} Expected flip probability for spins in an SSC of size $r^* N$, where $r^*$ is the cluster-size value at which the entropy is maximum. Inset: $P_r(h)$ for three different temperatures.}
\end{figure}

This behavior has a simple interpretation in terms of the effect of random fluctuations on in-cluster spins. Each spin's internal field $u_i$ is aligned with the total field $h_i$ and, since  fluctuations involving a finite number of spins do not result in macroscopic contributions to the difference between in- and out-cluster induced fields, the alignment between in-cluster and total fields is largely insensitive to fluctuations.
In other words, in-cluster spins provide a reinforcement mechanism to one another through the in-cluster dominated field that compensates for random fluctuations, which is absent for out-cluster spins.
When several SSC overlap, a competition between the influence of the different SSC forming islands of constrained spins, emerges.
%
%
SSC appear in sizes up to $r_0N$ depending on temperature as shown in the inset of Fig.~\ref{fig:entropy}, with $r_0$ approaching $1$ as $T$ decreases.
Because of numerical instabilities in solving saddle point equations, it is impossible evaluate precisely the temperature at which $r_0$ approaches $1$. This temperature is supposed to be in the range $[T_d,T_{TAP}]$, where in this model $T_{TAP} \sim 0.764$, as obtained in the SM.

The existence of SSC at high temperatures suggests that their existence does not trivially relate to slow-evolving spins, which do not exist in this regime.
Results obtained for different resilience parameter values $\epsilon$ (see Eq.~(\ref{eq:SSCcond})), shown in Fig.~\ref{fig:eps01}, exhibit a strong presence of low-field SSC at high temperatures. Firstly, we note that the absolute value of the local field experienced by internal spins is an increasing function of $\epsilon$. A slow-evolving spin is characterized by a strong field such that $\beta |h_i|$ is large. Thus, SSC can be associated with slow-evolving spins only if they exist for a sufficiently high $\epsilon$ and this argument suggests that $\epsilon$ should be an increasing function of $T$.
For instance, the scaling $|h| \sim \sqrt{\epsilon}$ leads to $\epsilon \sim T^2$.
Fig.~\ref{fig:eps01} shows that as we increase $\epsilon$, fewer and fewer SSC exist with a much stronger effect exhibited at high temperatures, as demonstrated in the inset. Even a very small value of $\epsilon$ (e.g. $\epsilon=0.1$) decimates the number of SSC at high temperature (e.g. $T\!=\!10$) while strong clusters are unaffected at lower temperatures (e.g. $T\!=\!1$).

To verify the analytical picture obtained, we compute the entropy of SSC by exhaustive search or sampling in small 3-spin systems. As shown in Fig.~\ref{fig:small}, SSC exist at high temperature and the entropy of large SSC increases when the temperature $T$ decreases in small systems, in agreement with the analytical results shown in Fig.~\ref{fig:entropy}. On the other hand, SSC with small $r$ values may be absent in small system sizes due to finite-size effects as shown in Fig.~\ref{fig:small}, compared to the thermodynamic limit in Fig.~\ref{fig:entropy}.

To gain a quantitative measure of how slow the in-cluster spins are, we computed the corresponding distribution of local fields $P_r(h)$ using Eqs.~(\ref{Prh})-(\ref{Prh2}). We employed this distribution to compute the expected flip probability spins in an SSC of size $r^{*}N$
\be
\pi_{\mathcal{E}}(T) = \int d h P_{r^*}(h) \frac{e^{-\beta |h|}}{2 \cosh (\beta h)}\:,
\ee
where $r^*$ is the the value at which $S_{\beta}(r)$ is maximum. The expected probability $\pi_{\mathcal{E}}(T)$ rapidly decreases to zero as $T$ decreases, as shown in Fig.~\ref{fig:Escape}. The inset shows $P_r^{*}(h)$ for different temperatures.

The above results show that when the temperature decreases, SSC in the 3-spin model increase in number, become more extensive in size, and are more stable against thermal fluctuations. Since these clusters are self-sustained and are not necessarily consistent with the equilibrium state of the system, their stable existence slows the system's dynamics towards equilibrium. In optimization problems, the presence of SSC would induce computational hardness since local algorithms will not escape states with SSC on the search for optimal solutions.

\emph{Summary:} We proposed a theoretical framework to address the issue of slow-evolving variables in spin systems at the microscopic level based on the concept of SSC, which can be viewed as regions of interdependent mutually-stabilizing spins. As the temperature decreases, strong SSC emerge and encompass increasingly larger fraction of the system with inevitable conflicts between competing clusters.  We provide new microscopic perspective on the dynamical slowing down observed in spin systems at low temperatures, complementing existing macroscopic understanding with the potential of providing new algorithmic optimization tools for hard computational problems through the destabilization of SSC in problems that can be mapped onto spin systems.

This work is supported by The Leverhulme Trust grant RPG-2013-48 and the Research Grants Council of Hong Kong (CHY, Grant No. 18304316). We are grateful to Leticia Cugliandolo for pointing us to references~\cite{barrat1999time, ricci2000glassy, roma2006signature, roma2010domain} and for helpful comments. We would also like to thank Pierfrancesco Urbani for interesting discussions.

\bibliographystyle{unsrt}

\providecommand{\noopsort}[1]{}\providecommand{\singleletter}[1]{#1}%

\newpage

\section{Supplemental Information}

In this section we first describe some aspects of the Ising $p$-spin model, and then outline the derivation of the Self-Sustained Clusters' (SSC) entropy.
Properties of the $p$-spin model are introduced by presenting the Franz-Parisi potential and the computation of the complexity of the system.
While the Franz-Parisi method is very interesting on its own right, it is also very instructive for our purposes because it allows one to discuss some useful technical issues related to the computation of the SSC entropy.
The computation of the complexity gives a different perspective on the emergence of TAP states in this model and provides an estimation of the temperature, $T_{TAP}$, at which such states appear.
The two approaches are independent and complementary. More aspects of the rich phenomenology of the $p$-spin model can be found in~\cite{mezard2009information, crisanti2005complexity, rizzo2013replica}. Here we restrict the description to the case of $p=3$.

\subsection{Franz-Parisi potential}
\begin{figure}[ht]
\includegraphics[width=160mm]{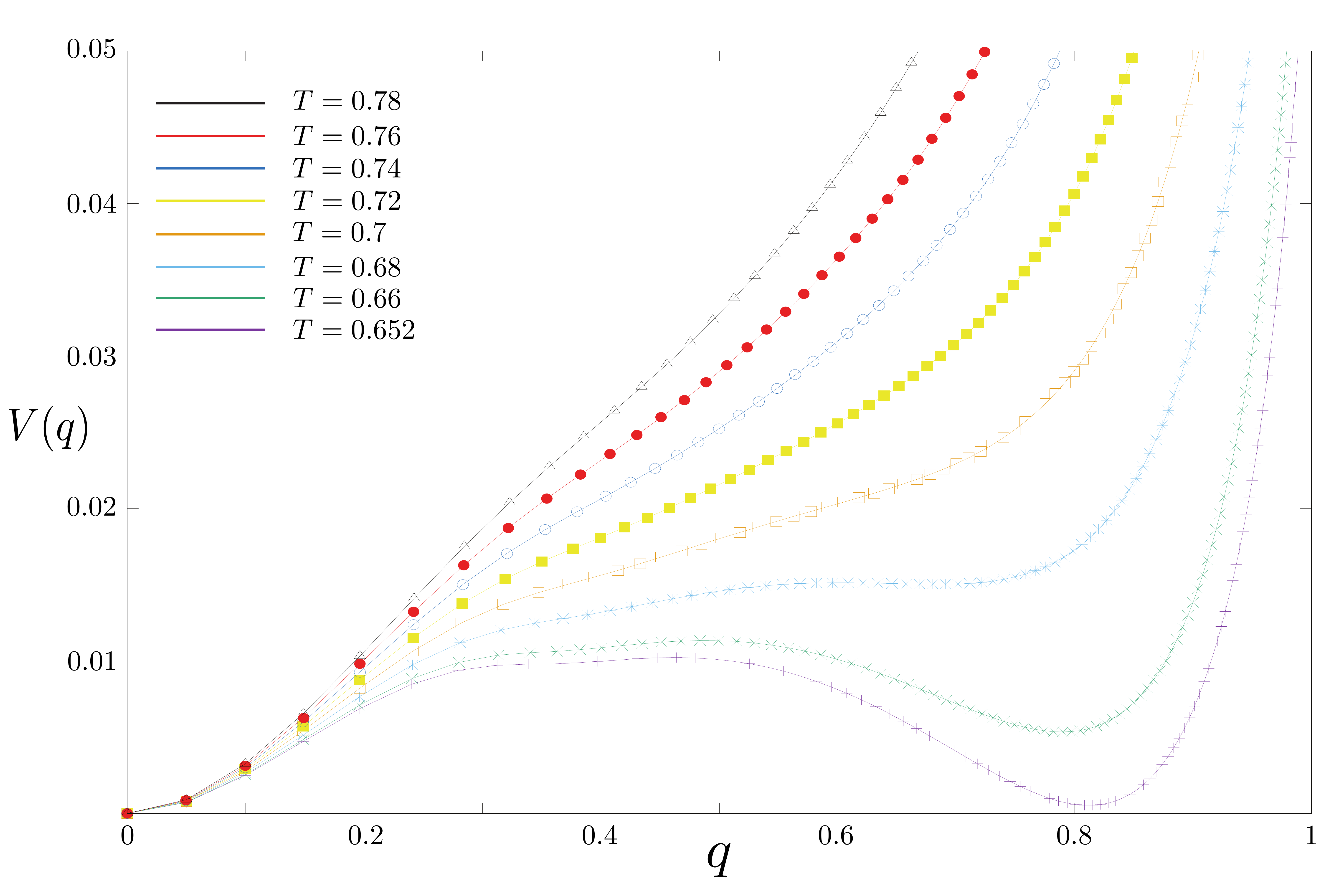}
\caption{\label{fig:FP} Franz-Parisi potential of the Ising 3-spin model. The potential has only one minimum at $q=0$ in the high temperature phase but as $T\rightarrow T_d \sim 0.681$, a metastable minimum appears at $q^*>0$. This second minimum becomes the global one for $T<T_K \sim 0.652$.}
\end{figure}

The Franz-Parisi potential~\cite{FranzParisi} is defined by introducing the free energy per spin of a system that is constrained to have an overlap $q$ with a reference configuration $\us$
\be
- N \beta F_{\us,\{J_{ijk}\}}(q) = \log \sum_{\ux} e^{-\beta H(\ux)} \delta \left(Nq - \sum_i \sigma_i s_i\right)\:.
\label{eq:FP1}
\ee
This quantity depends on the chosen configuration $\us$ and the interaction variables $\{J_{ijk}\}$. To obtain an expected value of (\ref{eq:FP1}) we assume self-averaging properties with respect to both and compute
\be
V(q) = \mathds{E}_J \mathds{E}_{\us} \left[ F_{\us,\{J_{ijk}\}}(q) - F \right]
\ee
where, as in the main text, $\mathds{E}_J$ denotes the average over disorder and $\mathds{E}_{\us}$ the average over the Boltzmann weight for configuration $\us$.
The Hamiltonian of the model is defined in Eq.~(\ref{eq:hamiltonianpspin}) and the quenched disorder is such that $J_{ijk}$ are i.i.d. Gaussian random variables, of mean $0$ and variance $p! / (2 N^{p-1})$, where $N$ is the number of spins.
The quantity $V(q)$ is the large deviation function of the probability to observe two configurations extracted from the equilibrium (Boltzmann) distribution with overlap $q$, i.e. the Parisi function $P(q)$. One of the main reasons to study $V(q)$ is that it contains information about the dynamical transition $T_d$ that is missing in both $P(q)$ and $F$.
In order to compute $V(q)$ we need to use the replica trick twice. This can by done invoking Eq.~(\ref{eq:eqrep2}) to calculate the Boltzmann weight, and the identity
\be
- N \beta F_{\us,\{J_{ijk}\}}(q) = \lim_{m \rightarrow 0} \partial_m \sum_{ \{\ux\} } e^{-\beta \sum_{\alpha=1}^m H\left[ \ux^{(\alpha)}\right]} \prod_{\alpha=1}^m \delta\left(Nq - \sum_i \sigma_i^{(\alpha)} s_i^{(1)}\right)\:,
\ee
where $\{\ux\}=\{\ux^{(1)}\ldots \ux^{(m)}\}$, to calculate with logarithm inside the averages.
It is useful to introduce the order parameters
\ber
\qss^{ab} & = \frac{1}{N} \sum_i s^{(a)}_i s^{(b)}_i  \label{eq:over1}\\
\qxx^{\alpha \beta} & = \frac{1}{N} \sum_i \sigma^{(\alpha)}_i \sigma^{(\beta)}_i \label{eq:over2}\\
Q_{s \sigma}^{a \alpha} & = \frac{1}{N} \sum_i s^{(a)}_i \sigma^{(\alpha)}_i  \label{eq:over3}\:,
\eer
and the corresponding conjugate order parameters (with hats) by using the integral representation of the Dirac delta function, illustrated here for $\qss$:
\be
1= \prod_{a<b} \int d \qss^{a b} \frac{d \hqss^{a b}}{2 \pi} e^{-i \hqss^{ab} \left( N \qss^{ab} - \sum_{i=1}^N s_i^{(a)}  s_i^{(b)} \right) }\:.
\label{eq:intrepdelta}
\ee
These manipulations lead to
\begin{widetext}
\be
-\beta V(q) =  \beta^2 \left( \frac{1}{4} -\frac{1}{4}\qxx^3 + \frac{1}{2}q^3 - \frac{1}{2}P^3 \right) + \frac{1}{2}( \qxx-1) \hqxx - \hat{q} q + \hat{P} P + \mathcal{I}\:,
\label{eq:v(q)FPprior}
\ee
where $\mathcal{I}  = \int D \uz \: h(z_1, z_2, \hat{q},\hat{P})$,
\be
h(z_1, z_2, \hat{q},\hat{P})  = \frac{ \log 2 \cosh\left(z_2 + (\hat{q}-\hat{P}) \right) e^{z_1} + \log 2 \cosh\left( z_2 - (\hat{q}-\hat{P}) \right) e^{-z_1} } {2 \cosh z_1}
\ee
and the Gaussian measure $D \uz$ is
\be
D\uz=\sqrt{\frac{\det \left( U \right)^{-1} }{ (2\pi)^2}} \prod_{k=1}^2 dz_k \exp \left \{ -\frac{1}{2} \uz^T  U^{-1}   \uz \right \}\:, \qquad  U = \left(
\begin{array}{cc}
\hqss & \hat{P}\\
\hat{P} & \hqxx \\
\end{array}
\label{eq:eqD}
\right)\:.
\ee
Equation (\ref{eq:v(q)FPprior}) is correct under the RS assumption that holds for $\forall \: T > T_K$~\cite{FranzParisi}.
The order parameters $\qss$ and $\qxx$ are the off diagonal terms of the matrices defined in Eqs.~(\ref{eq:over1}) and (\ref{eq:over2}). The delta function in Eq.~(\ref{eq:FP1}) sets $Q_{s \sigma}^{1\alpha}=q, \: \forall \alpha$, and since the first row of this matrix is $\alpha$ independent we also set $Q_{s \sigma}^{a \alpha}= P^a, \: a \neq 1, \forall \alpha$ and under the RS ansatz $P^a=P, \: a\neq 1$. Similarly, we set $\hat{Q}_{s \sigma}^{1 \alpha}=\hat{q}, \: \forall \alpha$ and
$\hat{Q}_{s \sigma}^{a \alpha}=\hat{P}, \: a \neq 1, \forall \alpha$. The saddle point equations set the other order parameters to
\ber
\hqss & = \frac{3}{2} Q^2_{ss} \beta^2 \:, \label{eq:hat1}\\
\hqxx & = \frac{3}{2} \qxx^2 \beta^2 \:, \label{eq:hat2}\\
\hat{P} & = \frac{3}{2} P^2 \beta^2\:, \label{eq:hat3}
\eer
while the original order parameters obey the equations
\be
\frac{1}{2}\qss - \frac{1}{2} + \int D \uz  \left [ \frac{1}{2}z_1^2 \hqss^{-2} - \frac{1}{2} \hqss^{-1}  \right ]  \log 2 \cosh z_1 = 0\:, \label{eq:SPFP1}
\ee
\be
\frac{1}{2}\qxx - \frac{1}{2} + \int D \uz  \left [  \frac{1}{2} \left( \sum_l [U^{-1}]_{2l} z_l \right)^2 - \frac{1}{2} \left[ U^{-1}\right]_{22} \right ]   h(z_1, z_2, \hat{q},\hat{P})   = 0	\:, \label{eq:SPFP2}
\ee
\be
P + \int D \uz  \left [  \sum_{lp} [U^{-1}]_{1l} [U^{-1}]_{2p}z_l  z_p  - \left[ U^{-1}\right]_{12} \right ]   h(z_1, z_2, \hat{q},\hat{P})   = 0 \:, \label{eq:SPFP3}
\ee
\be
q = \int D \uz \left[ \tanh \left(z_2 + (\hat{q}-\hat{P}) \right) e^{z_1} - \tanh \left(z_2 - (\hat{q}-\hat{P}) \right) e^{-z_1} \right] \left[2 \cosh z_1\right]^{-1} \label{eq:SPFP4}\:.
\ee
Equation~(\ref{eq:SPFP1}) is independent of $P$, $\qxx$ and their conjugate order parameters because the configuration related variables ($\us$ system) are unaffected by the computation of the constrained free energy in Eq.~(\ref{eq:FP1}). Notice that the overlap value $q$ in Eq.~(\ref{eq:SPFP4}) does not require to be optimized.
Before solving these equations, we notice that as long as $T> T_K$, $\qss=\hqss=0$ and this condition leads to $P=\hat{P}=0$. Thus, rather than solving a system of equations, we end up with solving the single equation,
\be
 \hqxx = \frac{1}{2} \int Dt \left[ \tanh^2 \left(\hat{q}-\sqrt{\hqxx}t\right) + \tanh^2 \left(\hat{q}+\sqrt{\hqxx}t\right) \right]\:
 \label{eq:FPhqxxpost}
 \ee
and compute the corresponding value for $q$, given by
\be
q = \frac{1}{2} \int Dt \left[  \tanh \left(\hat{q}-\sqrt{\hqxx}t\right) + \tanh \left(\hat{q}+\sqrt{\hqxx}t\right) \right]~,
\label{eq:PFoverqSPeq}
\ee
where $\hqxx$ is still given by Eq.~(\ref{eq:hat2}) and $Dt = \mathcal{N}(0,1)$ is a Normal Gaussian distribution of zero mean and variance $1$.  Moreover, thanks to this simplification, Eq.~(\ref{eq:v(q)FPprior}) becomes
\be
-\beta V(q) = \beta^2 \left( \frac{1}{4} -\frac{1}{4}\qxx^3 + \frac{1}{2}q^3 \right) + \frac{1}{2}( \qxx-1) \hqxx - \hat{q} q + \mathcal{I}'
\label{eq:v(q)FPpost}
\ee
where
\be
\mathcal{I}' = \frac{1}{2} \int Dt \left[ \log 2 \cosh \left(\hat{q} + \sqrt{\hqxx} t \right) + \log 2 \cosh \left(\hat{q} - \sqrt{\hqxx} t \right) \right]\:.
\ee
\end{widetext}
We can solve Eq.~(\ref{eq:FPhqxxpost}) for each $\{\beta, \hat{q}\}$ iteratively. Finally, plugging the solution in Eqs.~(\ref{eq:v(q)FPpost}) and (\ref{eq:PFoverqSPeq}) we obtain a value for $V(q)$. The results can be seen in Fig.~(\ref{fig:FP}).
The dynamical temperature $T_d$ is defined as the temperature at which the potential develops a local minimum at an overlap value $q^*$ and the Kauzmann temperature $T_K$ as the temperature at which the local minimum becomes a global one.
We see that $T_d\sim 0.681$ and $T_K\sim 0.652$.

\subsection{Complexity}
\begin{figure}[ht]
\includegraphics[width=160mm]{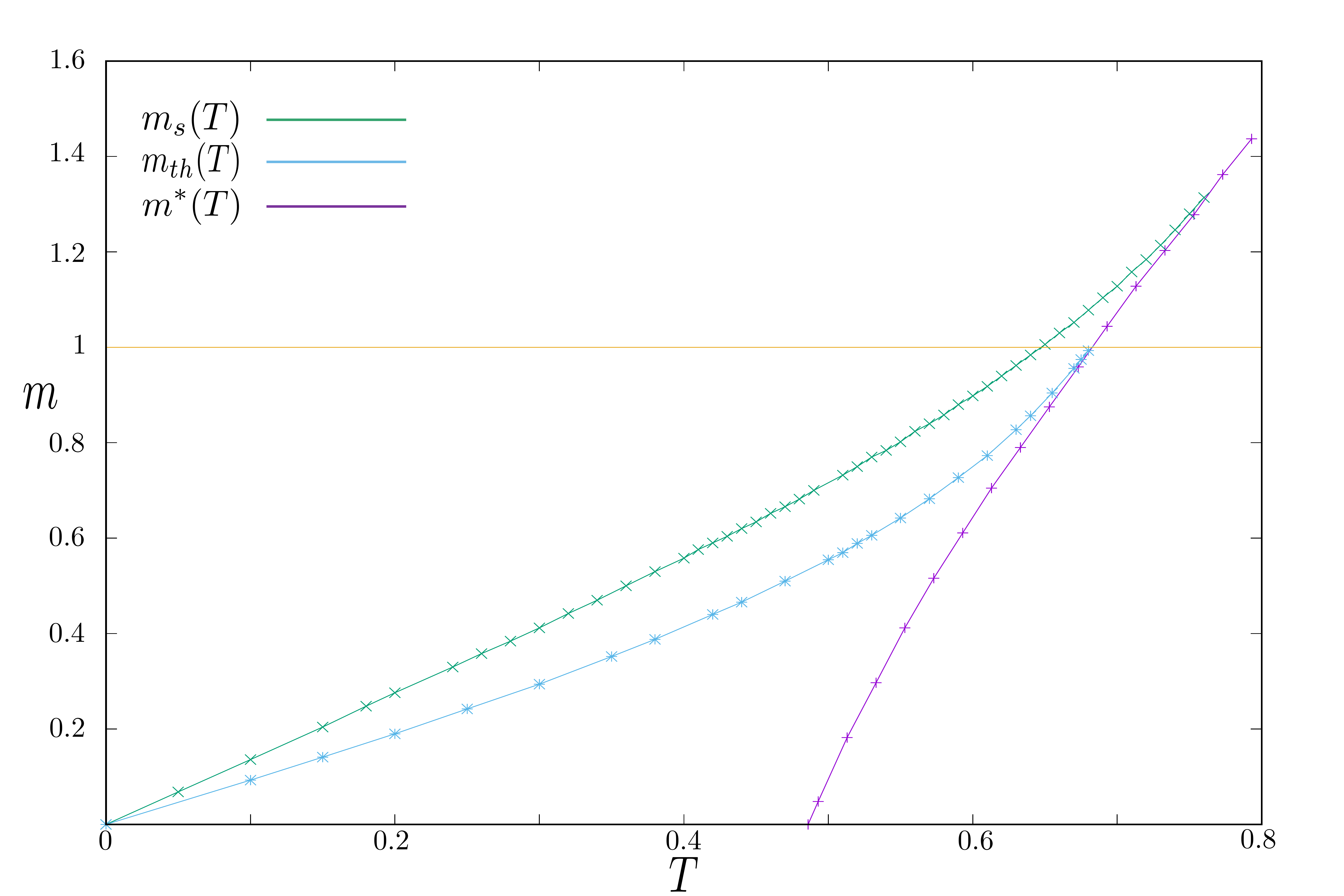}
\caption{\label{fig:m-Tphasediag} Phase diagram of the Ising $3-$spin in the $m-T$ plane. The Kauzmann temperature $T_K\sim0.652$ is defined by the intersection between $m=1$ and $m_s(T)$, and the dynamical temperature $T_d\sim0.681$ by the intersection between $m=1$ and the line $m^*(T)$. The lines $m_{th}(T)$ and $m^*(T)$ merge for $T>T_d$. The TAP temperature $T_{TAP}\sim 0.764$ is defined by the intersection between $m_s(T)$ and $m_{th(T)}$. }
\end{figure}

Computing the complexity of the $p$-spin model offers a different perspective.
The name complexity denotes the entropy of the number of metastable states that dominate the Boltzmann weight in the dynamical phase.
It can be computed by solving the TAP equations~\cite{thouless1977solution} of the model~\cite{crisanti2005complexity}, or by counting the number of pure states (or TAP states) of the system.
A detailed description of pure states can be found in~\cite{zamponiRev}. They can be defined as measures on the configuration space with vanishing connected correlation functions between distant degrees of freedom.
This notion is intuitively related to equilibrium states, where the response function vanishes at long distance. Since in mean field model there is no notion of distance, pure states are factorized such that
\be
P^{\omega}(s_1, \ldots, s_N) = \prod_{i=1}^N p^{\omega}(s_i) \:, \qquad p^{\omega}(s_i) = \frac{1+s_i m_i^{\omega}}{2}
\ee
where $\omega$ identifies one of these states. In other words, in mean field models pure states are such that all the connected correlation functions vanish.
At high temperatures only one such state exist, the paramagnetic state, where $m_i=0, \forall i$.
As $T$ decreases, new states emerge. In this situation, the partition function can be decomposed as follows
\be
Z = \sum_{\us} e^{-\beta H(\us)} = \sum_{\omega} e^{-\beta N f_{\omega}}\:,
\ee
where $f_{\omega}$ denotes the internal free energy of the state $\omega$.
In models without quenched disorder, when several pure states exist they can be selected by introducing an external vanishing field. Unfortunately, spin glass models do not allow for a similar procedure because it is an unfeasible to slightly perturb, locally, each spin in the correct direction.
In order to solve this, Monasson~\cite{monasson1995structural} introduced the method of coupled replica, guiding each
other they all end up in the same pure state. Thus the partition function of $m$ coupled replica (referred to as $m$-system from now on) can be written as
\be
Z_m = \sum_{\omega} e^{-\beta m N f_{\omega}} = \int_{f_m}^{f_{th}} df  \: \sum_{\omega} \delta (f-f_{\omega}) e^{-\beta m N f} = \int_{f_m}^{f_{th}} df \: e^{N [ -\beta m f + \Sigma(f) ] }
\ee
where $f_m$ and $f_{th}$ define the limits where pure states can be found and $\Sigma(f(m,T))$ is, by definition, the complexity of the $m-$system. Eventually, we will be interested in the limit $m\rightarrow 1$. This integral can be computed using the saddle point method and gives
\be
\Phi(m,T) = -\frac{T}{N} \mathbf{E}_J \log Z_m = m f^* (m,T) - T \Sigma (f^*(m,T))\:.
\ee
$f^*(m,T)$ is the free energy of the pure states of the $m$-system that dominates the Boltzmann weight at temperature $T$. Both $f^*(m,T)$ and $\Sigma(f^*(m,T))$ can be found by differentiating $\Phi(m,T)$:
\be
\Sigma(f^*(m,T)) = m^2 \frac{\partial \left[ m^{-1} \beta \Phi(m,T) \right]}{\partial m}
\label{eq:sigma1rsb}
\ee
\be
f^*(m,T) = \frac{\partial \Phi(m,T)}{\partial m}~;
\label{eq:freeenergy1rsb}
\ee
thus, $m$ can be used as a dummy variable to compute numerically $\Sigma(f)$, which is the entropy of the pure states with free energy equal to $f$.
This is a general protocol that can be carried out in every model. All we need to do is compute $\Phi(m,T)$ for which we use the replica method:
\be
\Phi(m,T) =  -\frac{T}{N} \mathbf{E}_J \log Z_m = -\frac{T}{N} \lim_{n \rightarrow 0} \partial_n \mathbf{E}_J  (Z_m)^n
\ee
Introducing the integral representation of~(\ref{eq:intrepdelta}) we obtain the order parameters $Q^{ab}$ and $\hat{Q}^{ab}$ with indices in the $m \times n$ dimensional space.
These matrices contain $n$ groups of $m$-coupled replicas: it is thus natural to employ a $1-$RSB ansatz where the off-diagonal elements indexed by $\{ a,b \}$ are zero for $a,b$ not in the same block, and take a positive value for $\{ a,b \}$ in the same block. This manipulation leads to
\be
\Phi(m,\hat{Q}_{ss},Q_{ss},T) = -T m \left[ -\frac{(m-1)}{2} \hat{Q}_{ss} Q_{ss} + \frac{\beta^2}{4} \left[ (m-1) Q_{ss}^3 + 1 \right] + m^{-1}\phi - \frac{1}{2} \hat{Q}_{ss} \right ]
\label{eq:1rsbevalphi}
\ee
where
\be
\phi=\log \int Dt \left[ 2\cosh(\sqrt{Q_{ss}} t)\right]^m
\ee
and $\hat{Q}_{ss}$ and $Q_{ss}$ are linked through the saddle point equations
\be
Q_{ss} = \frac{2}{m-1} \left \{ - \frac{1}{2} + \frac{\sqrt{\hat{Q}^{-1}_{ss}}}{2}  \frac{ \displaystyle \int Dt \: t \left[ 2 \cosh \sqrt{\hat{Q}_{ss}} t \right]^m \tanh \left( \sqrt{\hat{Q}_{ss}} t \right) }{\displaystyle \int  Dt \left[ 2 \cosh \sqrt{\hat{Q}_{ss}} t \right]^m  } \right \}
\label{eq:eqQssMon}
\ee
with $\hat{Q}_{ss}$ given by Eq.~(\ref{eq:hat1}). Equation~(\ref{eq:eqQssMon}) has three solutions but we are interested only in the largest one. This is because, at a given $m>1$ and $T$, the function $\Phi(m,\hat{Q}_{ss}=3 \beta^2 Q^2_{ss}/2,Q_{ss},T)$ has three stationary points as a function of $Q_{ss}$, the smallest one ($Q_{ss}=0$) and the largest one ($Q_{ss}=q^*$) being minima and the intermediate one being a maximum.
These two values correspond to the overlap of different replica in the $m$-system. As $m\rightarrow 1^{+}$ the two minima are degenerate but, because of the non-zero coupling between replica, $q^*$ has to be preferred.
Using Eq.~(\ref{eq:1rsbevalphi}) in Eqs.~(\ref{eq:sigma1rsb}) and~(\ref{eq:freeenergy1rsb}) we finally obtain
\be
\Sigma(f^*(m,T)) = m^2 \left[ \frac{\hat{Q}_{ss} Q_{ss}}{2} - \frac{\beta^2}{4} Q_{ss}^3 + m^{-2} \phi - m^{-1} \partial_m \phi \right ]
\ee
\be
f^*(m,T) = (-T) \left[ -\frac{m-1}{2} \hat{Q}_{ss} Q_{ss} + \frac{\beta^2}{4}   \left[ (m-1) Q_{ss}^3 + 1 \right] + m^{-1}\phi - \frac{1}{2} \hat{Q}_{ss} \right] + \frac{T}{m}\Sigma(f^*(m,T))
\ee
that can be used to compute $\Sigma(f)$ at different temperatures by using $m$ as a parameter.
This approach provides a phase diagram in the $m-T$ plane shown in Fig. \ref{fig:m-Tphasediag}, where three lines can be identified:
$m^*(T)$ is the line below which Eq.~(\ref{eq:eqQssMon}) has no solution with $q^*>0$. Below this line, the system in the paramagnetic state and the intersection between this line and that of $m=1$ identifies the dynamical transition.
The line where the complexity vanishes is defined as $m_s(T)$. Its intersection with the line $m=1$ identifies $T_K$.
The intermediate line, denoted as $m_{th}(T)$ defines a non-physical region in this plane, bounded between $m_{th}(T)$ and $m^*(T)$. 
The line
$m_{th}(T)$ crosses and merges with $m^*(T)$ for $T=T_d$.
This behavior can be understood by observing that a large $m$ has the same effect of a small $T$. Thus, at a given $T$, the region where $f(m,T)$ and $\Sigma(f(m,T))$ grow with $m$ is un-physical (since both are expected to decrease as $T$ decreases).
This argument also suggests that $m$ can be effectively used to probe non-equilibrium TAP-states. In fact, by definition, in the region between $m_{th}(T)$ and $m_s(T)$ the complexity is positive and we see that this region extends for $T>T_d$, where the original system (that can be recovered for $m\rightarrow1$) is in the paramagnetic phase.
The complexity shrinks to zero as $T$ grows and the temperature at which $m_s(T)$ and $m_{th}(T)$ merge is denoted as $T_{TAP}$, in which TAP states exist as non-equilibrium states; for the 3-spin model $T_{TAP} \sim 0.764$.

\subsection{Entropy of Self-Sustained Clusters}
Similarly to the computation of the Franz-Parisi potential described above, replicas are introduced to calculate the logarithm in Eq.~(\ref{eq:eqSrs}) and the Boltzmann weight in Eq.~(\ref{eq:eqS}).
Moreover, in order to be consistent with the notation used in the main text and with the computation of the Franz-Parisi potential, we denote by $\sigma$ the ``internal" variables used to define clusters, and by $s$ the ``external" variables referring to the spin configuration $\us$ drawn from the Boltzmann weight.
To deal with the SSC membership condition, stated in Eq.~(\ref{eq:SSCcond}) we use the definition of the local field~(\ref{eq:eqheff}) and introduce the following two replicated fields
\ber
h_i  = && \frac{1}{2} \sum_{j ,k } J_{ijk} s_j s_k  \:, \label{eq:fieldhSI} \\
\eta^{(\alpha)}_i=&&\frac{1}{2}\sum_{j , k } J_{ijk} \: s^{(1)}_j s^{(1)}_k \sigma_j^{(\alpha)} \sigma_k^{(\alpha)}\:, \label{eq:fieldeta} \\
\mu^{(\alpha)}_i=&&\sum_{j , k } J_{ijk} \: s^{(1)}_j s^{(1)}_k  \sigma_j^{(\alpha)}  \label{eq:fieldmu} \:.
\eer
We define the entropy of SSC $\mathcal{S}_{\beta}(r)$ by
\be
\mathcal{S}_{\beta}(r) = \mathds{E}_J \mathds{E}_{\us} \left[ \: \mathcal{S}(r\:| \us,\{J_{ijk}\})\right]\:,
\label{eq:eqSSI}
\ee
where $\mathcal{S}(r\:| \us,\{J_{ijk}\})$ denotes the entropy of SSC in a given configuration $\us$ and quenched disorder realization $\{J_{ijk}\}$,
\be
\mathcal{S}(r\:| \us,\{J_{ijk}\})=N^{-1} \: \log \sum_{\ux} \mathds{I}_{\ux}(\us) \: \delta \left(r N - \sum_{i=1}^N \frac{1+\sigma_i}{2}\right)\:.
\label{eq:eqSrsSI}
\ee
Employing the integral representation of the delta function, used to enforce the definitions of Eqs.~(\ref{eq:fieldhSI}), (\ref{eq:fieldeta}) and (\ref{eq:fieldmu}),
and introducing the notation
\be
\mathcal{D}\mathcal{B}\hat{\mathcal{B}} = \prod_{i} \frac{d\hat{h}_i d h_i}{2 \pi}  \prod_{i,\alpha} \frac{d\hat{\eta}^{(\alpha)}_i d \eta^{(\alpha)}_i}{2 \pi} \prod_{i,\alpha} \frac{d\hat{\mu}^{(\alpha)}_i d \mu^{(\alpha)}_i}{2 \pi}\:,
\ee
Eq.~(\ref{eq:eqSSI}) can be written as
\be
N \mathcal{S}_{\beta}(r) = \lim_{n \rightarrow 0} \lim_{m \rightarrow 0} \mathds{E}_J \left[ \partial_m \mathcal{N}_{nm} \right]
\label{eq:eqdefEntr}
\ee
where the replicated quantity $\mathcal{N}_{nm}$ is given by
\ber
\mathcal{N}_{nm} && =  \sum_{\{\us\}} \sum_{\{\ux\}} \int \mathcal{D}\mathcal{B}\hat{\mathcal{B}} \exp \left \{
-i \sum_i \hat{h}_i h_i -
i \sum_{\alpha, i} \hat{\eta}_i^{(\alpha)} \eta_i^{(\alpha)} -
i \sum_{\alpha, i} \hat{\mu}_i^{(\alpha)} \mu_i^{(\alpha)}
\right \}      \nonumber \\
&& \times  \exp \left \{
+\sum_{\alpha}  \sum_{ijk} \frac{i \hat{\eta}^{(\alpha)}_i }{2}   J_{ijk}s^{(1)}_j s^{(1)}_k \sigma^{(\alpha)}_j \sigma^{(\alpha)}_k
+\sum_{\alpha}  \sum_{ijk}  i \hat{\mu}^{(\alpha)}_i    J_{ijk}s^{(1)}_j s^{(1)}_k \sigma^{(\alpha)}_j
\right\}  \nonumber \\
&& \times \exp \left \{  \sum_{ijk} \frac{i  \hat{h}_i }{2} J_{ijk} s^{(1)}_j s^{(1)}_k \right \}   \exp  \left [ -\beta \sum_{a=1}^n H\left(\us^{(a)}\right) \right ] \prod_{\alpha=1}^m \mathds{I}^r_{\ux^{(\alpha)}}\left(\us^{(1)}\right) \label{eq:eqNnm1} \:,
\eer
the indicator function $ \mathds{I}^r_{\ux^{(\alpha)}}\left(\us^{(1)}\right) $ is defined by
\be
\mathds{I}^r_{\ux^{(\alpha)}}(\us^{(1)}) = \prod_{i=1}^{N} \left \{ \frac{1-\sigma^{(\alpha)}_i}{2} + \frac{1+\sigma^{(\alpha)}_i}{2} \theta \left[  \left( \mu_i^{(\alpha)} + \eta^{(\alpha)}_i - h_i \right) h_i - \epsilon  \right] \right \} \delta \left(r N - \sum_{i=1}^N \frac{1+\sigma^{(\alpha)}_i}{2}\right)\:,
\label{eq:eqIrep}
\ee
and the short-hand notation $\{\us\}$, $\{\ux\}$ stands for $\{\us^{(1)}\ldots \us^{(n)}\}$, $\{\ux^{(1)}\ldots \uxÄ^{(m)}\}$.
The term $h_i$ contained in the argument of the Heaviside function in~(\ref{eq:eqIrep}) does not have any replica index because it depends only on the configuration $\us=\us^{(1)}$.
The effective fields $h_i$, $\eta_i^{(\alpha)}$ and $\mu_i^{(\alpha)}$ are used to make the argument of the Heaviside function disorder independent, so that the quenched average can be performed more easily.
This average leads to the definition of several order parameters listed in Table \ref{table:tableop}.
\begin{table}[!ht]
\centering
\caption{Order parameters introduced in the computation of the entropy of SSC. We also make use of the order parameters defined in Eqs.~(\ref{eq:over1}) and~(\ref{eq:over2}). }
\begin{tabular}{ l  l  l }
\hline
$\qeer =-\frac{1}{N}\sum_i  \hat{\eta}_i^{(\alpha)} \hat{\eta}_i^{(\beta)}\:$
& $\: \qmmr=-\frac{1}{N}\sum_i \hat{\mu}_i^{(\alpha)} \hat{\mu}_i^{(\beta)} \:$
& $\: \qemr = -\frac{1}{N}\sum_i  \hat{\eta}_i^{(\alpha)}  \hat{\mu}_i^{(\beta)}\:$ \\
$ \chsxr=\frac{i}{N}\sum_i  \hat{h}_i s_i^{(1)}\sigma_i^{(\alpha)} \:$
& $\: \qesxr=\frac{i}{N}\sum_i  \hat{\eta}_i^{(\alpha)} s_i^{(1)} \sigma_i^{(\beta)}\:$
&  $\: \qmsxr=\frac{i}{N}\sum_i \hat{\mu}_i^{(\alpha)} s_i^{(1)} \sigma_i^{(\beta)}\:$  \\
$ \qhsr=\frac{i}{N}\sum_i  \hat{h}_i s_i^{(a)}\:$
& $\: \qesr=\frac{i}{N}\sum_i  \hat{\eta}_i^{(\alpha)} s_i^{(a)}\:$
&  $\: \qmsr=\frac{i}{N}\sum_i \hat{\mu}_i^{(\alpha)} s_i^{(a)}\:$  \\
$\chh = -\frac{1}{N}\sum_i  \hat{h}_i \hat{h}_i$
& $\: \cher =  -\frac{1}{N}\sum_i  \hat{h}_i \hat{\eta}^{(\alpha)}_i$
&  $\: \chmr =  -\frac{1}{N}\sum_i  \hat{h}_i \hat{\mu}^{(\alpha)}_i$ \\
$\mx = \frac{1}{N} \sum_i \sigma_i^{(\alpha)}$
& $\: \qssxr = \frac{1}{N} \sum_i s_i s_i^{a} \sigma_i^{\alpha} $
& \\
\hline
\end{tabular}
\label{table:tableop}
\end{table}
We also need to use the order parameters defined in Eqs.~(\ref{eq:over1}) and~(\ref{eq:over2}). The last order parameter $\qssxr$ is equal to $\mx$ for $a=1$, so we need to define it only for $a\neq1$.
Similarly to what has been done in Eq.~(\ref{eq:intrepdelta}), introducing the conjugate order parameters we obtain
\ber
 \mathds{E}_J \left[  \mathcal{N}_{nm} \right]   = && \int \mathcal{D} \mathcal{B}\hat{\mathcal{B}}  \exp \left \{
i \sum_i \hat{h}_i h_i +
i \sum_{\alpha, i} \hat{\eta}_i^{(\alpha)} \eta_i^{(\alpha)} +
i \sum_{\alpha, i} \hat{\mu}_i^{(\alpha)} \mu_i^{(\alpha)}
\right \} \prod_{\alpha \leq \beta }\frac{d \hqeer d \qeer} {2 \pi}  \nonumber  \\
&&
 \qquad  \prod_{\alpha \leq \beta }\frac{d \hqmmr d \qmmr} {2 \pi}
 \prod_{\alpha \beta }\frac{d \hqemr d \qemr} {2 \pi}
  \prod_{\alpha}\frac{d \hchsxr d \chsxr} {2 \pi}
 \prod_{\alpha \beta} \frac{d \hqesxr d \qesxr} {2 \pi}
  \nonumber \\
  &&
  \qquad \prod_{\alpha \beta} \frac{d \hqmsxr d \qmsxr} {2 \pi}
   \prod_{a}\frac{d \hqhsr d \qhsr} {2 \pi}
 \prod_{a \alpha} \frac{d \hqesr d \qesr} {2 \pi}
  \prod_{a \alpha} \frac{d \hqmsr d \qmsr} {2 \pi} \nonumber \\
 &&
 \qquad \prod_{\alpha < \beta }\frac{d \hqxxr d \qxxr} {2 \pi}
  \prod_{a < b }\frac{d \hqssr d \qssr} {2 \pi}
   \prod_{\alpha}\frac{d \hmxr d \mxr} {2 \pi}
\prod_{a\neq 1 \alpha} \frac{d \hqssxr d \qssxr} {2 \pi} \nonumber \\
  &&\qquad \frac{d \hchh d \chh} {2 \pi}
 \prod_{\alpha}\frac{d \hcher d \cher} {2 \pi}
 \prod_{\alpha}\frac{d \hchmr d \chmr} {2 \pi}  \nonumber \\
 && \qquad \exp \left[ N ( - \Psi( \{ Q, \hat{Q} \} ) + \Omega ( \{ Q \}) ) \right]  \nonumber \\
 && \qquad \sum_{\{\us\}} \sum_{\{\ux\}} \prod_{\alpha=1}^m
\mathds{I}^r_{\ux^{(\alpha)}}\left(\us^{(1)}\right)
\prod_i \exp \left [ N \Phi_{\{s_i\}, \{\sigma_i\} } \left( \hat{\mathcal{B}},  \{ \hat{Q}\}\right) \right ]
\eer
where the auxiliary function $\Psi( \{ Q, \hat{Q} \} )$, $\Omega ( \{ Q \}) )$ and $\Phi_{\{s_i\}, \{\sigma_i\} } \left( \hat{\mathcal{B}},  \{ \hat{Q}\}\right)$ are defined by:
\ber
 - i\Psi( \{ Q, \hat{Q} \} )=  &&  \left[ \sum_{\alpha \leq \beta} \hqeer \qeer + \sum_{\alpha \leq \beta} \hqmmr \qmmr
+\sum_{\alpha \beta} \hqemr \qemr + \sum_{\alpha < \beta} \hqxxr \qxxr +   \right. \nonumber  \\
&& + \sum_{a < b} \hqssr \qssr+\sum_{\alpha} \hcher \cher +\sum_{\alpha} \hchmr \chmr  + \sum_{\alpha} \hmxr \mxr
+ \nonumber   \\
&& + \sum_{\alpha} \hchsxr \chsxr + \sum_{\alpha \beta} \hqesxr \qesxr + \sum_{\alpha \beta} \hqmsxr \qmsxr  + \hchh \chh + \nonumber \\
&&\left. + \sum_{a} \hqhsr \qhsr + \sum_{a \alpha} \hqesr \qesr + \sum_{a \alpha} \hqmsr \qmsr + \sum_{a\neq 1 \alpha} \hqssxr \qssxr \right] \:,
\eer

\ber
\Omega ( \{ Q \}) = &&  3 \chh + 6 (Q_{\hat{h}s}^{11})^2 +\sum_{\alpha \beta} \left( 3 \qeer \left(\qxxr \right)^2  + 6 (\qesxr)^2 \qxxr \right) + \nonumber \\
&& + \sum_{\alpha \beta} \left( 3 \qmmr \left(2 \qxxr + 2 m_{\sigma}^{\alpha} m_{\sigma}^{\beta} \right) + 6 (\qmsxr)^2 + 12 Q_{\hat{\mu}s}^{\alpha 1} Q_{\hat{\mu}s\sigma}^{\beta 1 \alpha} m_{\sigma}^{\beta} + 6 Q_{\hat{\mu} s}^{\alpha 1}  Q_{\hat{\mu} s}^{\beta 1} \qxxr \right) + \nonumber \\
&& + \sum_{\alpha} \left( 6 \cher (\mxr)^2 + 12 \chsxr Q_{\hat{\eta} s}^{\alpha 1} \mxr + 12 \chmr \mxr + 12 \chsxr Q_{\hat{\mu} s}^{\alpha 1} + 12 Q_{\hat{h} s}^{11} Q_{\hat{\mu} s}^{\alpha 1} \mxr \right) + \nonumber \\
&& + \sum_{\alpha \beta} \left( 12 \qemr \qxxr \mxr + 12 \qesxr \qmsxr \mxr + 12 Q_{\hat{\eta} s}^{\alpha 1} \qmsxr \qxxr \right) + \nonumber \\
&& + \frac{1}{4} \left( \sum_{ab} (\qssr)^3  \beta^2 + 2 \beta \left( 3 Q_{\hat{s} s}^{11} + 3 \sum_{a\neq 1 } \qhsr (\qssr)^2 + 3 \sum_{\alpha} Q_{\hat{\eta} s}^{\alpha 1} (\mxr)^2 +          \right.  \right. \nonumber \\
&& \left. \left. + 3 \sum_{\alpha} \sum_{a\neq 1} \qesr (\qssxr)^2 + 6 \sum_{\alpha} Q_{\hat{\mu} s}^{\alpha 1} \mxr + 6 \sum_{\alpha} \sum_{a\neq 1} \qmsr \qssxr Q_{ss}^{1 a}  \right) \right)\:,
\eer

\ber
i \Phi_{\{s_i\}, \{\sigma_i\} } \left( \hat{\mathcal{B}},  \{ \hat{Q}\} \right) = && \left[
\sum_{\alpha \leq \beta} \hqeer \hat{\eta}_i^{(\alpha)} \hat{\eta}_i^{(\beta)}+
\sum_{\alpha \leq \beta} \hqmmr \hat{\mu}_i^{(\alpha)} \hat{\mu}_i^{(\beta)} + \sum_{\alpha \beta} \hqemr \hat{\eta}_i^{(\alpha)} \hat{\mu}_i^{(\beta)}   \right.  \nonumber \\
&&
- \sum_{\alpha < \beta} \hqxxr \sigma_i^{(\alpha)} \sigma_i^{(\beta)}
- \sum_{a < b} \hqssr s_i^{(a)} s_i^{(b)} +
\sum_{\alpha} \hcher \hat{h}_i \hat{\eta}_i^{(\alpha)}
+ \sum_{\alpha} \hchmr \hat{h}_i \hat{\mu}_i^{(\alpha)} \nonumber \\
&& \left. - \sum_{\alpha} \hmxr \sigma_i^{(\alpha)}  -i \sum_{a} \hqhsr \hat{h}_i s_i^{(a)}
-i \sum_{a \alpha} \hqesr \hat{\eta}_i^{(\alpha)} s_i^{(a)} -i \sum_{a \alpha} \hqmsr \hat{\mu}_i^{(\alpha)} s_i^{(a)}
 \right.  \nonumber \\
 && -\sum_{a \neq 1\alpha} \hqssxr s_i^{(1)} s_i^{(a)} \sigma_i^{(\alpha)}
 + \hchh \hat{h}_i \hat{h}_i - i \sum_{\alpha \beta} \hqesxr \hat{\eta}_i^{(\alpha)} s_i^{(1)} \sigma_i^{(\beta)}\nonumber \\
 && - \left. i \sum_{\alpha \beta} \hqmsxr \hat{\mu}_i^{(\alpha)} s_i^{(1)} \sigma_i^{(\beta)} -  i \sum_{\alpha} \hchsxr \hat{h}_i s_i^{(1)} \sigma_i^{(\alpha)} \right]\:.
 \label{eq:Phireplicated}
\eer
At this point we adopt the RS ansatz illustrated in Table \ref{table:RS1} and Table \ref{table:RS2}.
As mentioned in the main text, one should invoke a more complex hierarchical ansatz~\cite{mezard1990spin} when averaging over configurations $\us$ but the RS ansatz is valid for all temperatures higher than $T_K$ ~\cite{FranzParisi}. In fact, in this regime, the paramagnetic state is replaced by an exponential number of states whose overlap is zero~\cite{kirkpatrick1987dynamics, kurchan1993barriers, crisanti1995thouless}, from which a trivial Parisi function $P(q)=\delta(q)$ is obtained.
It means the overlap among two configurations, sampled from the equilibrium distribution, is zero because they are very likely to belong to different TAP states.
Our results are thus valid as long as $T>T_K$. We also employ an RS ansatz for the $\sigma$-related order parameters; since $\sigma$ variables are just labels used to define clusters, there is no obvious reason why a more complicated scheme should be invoked.
\begin{table}[!ht]
\centering
\caption{ }
\begin{tabular}{ l  l  }
\hline
\multicolumn{2}{c}{$\forall \alpha$}\\
\hline
$\hmxr=\hmx$    &   $\mxr=\mx$ \\
$\hcher=\hche$ &   $\cher=\che$ \\
$\hchmr=\hchm$ & $\chmr=\chm$ \\
$\hchsxr = \hchsx$ & $\chsxr = \chsx$ \\
\hline
\end{tabular}
\label{table:RS1}
\end{table}
\begin{table}[!ht]
\centering
\caption{ }
\begin{tabular}{ l  l  l  l }
\hline
\multicolumn{2}{c}{$\alpha \neq \beta$}&\multicolumn{2}{c}{$\alpha = \beta$}\\
\hline
$\hqeer=\hqee$ & $\qeer=\qee$ & $\hqeer=\hcee$ & $\qeer=\cee$ \\
$\hqmmr=\hqmm$ & $\qmmr=\qmm$ & $\hqmmr=\hcmm$ & $\qmmr=\cmm$ \\
$\hqemr=\hqem$ & $\qemr=\qem$ &  $\hqemr=\hcem$ & $\qemr=\cem$ \\
$\hqxxr=\hqxx$ & $\qxxr=\qxx$ & & \\
$\hqesxr=\hqesx$ & $\qesxr=\qesx$ &  $\hqesxr=\hcesx$ & $\qesxr=\cesx$ \\
$\hqmsxr=\hqmsx$ & $\qmsxr=\qmsx$ &  $\hqmsxr=\hcmsx$ & $\qmsxr=\cmsx$ \\
\hline
\multicolumn{2}{c}{$a \neq 1$}&\multicolumn{2}{c}{$a = 1$}\\
\hline
$\hqhsr=\hqhs$ & $\qhsr=\qhs$ & $\hqhsr=\hchs$ & $\qhsr=\chs$ \\
$\hqesr=\hqes$ & $\qesr=\qes$ & $\hqesr=\hces$ & $\qesr=\ces$ \\
$\hqmsr=\hqms$ & $\qmsr=\qms$ & $\hqmsr=\hcms$ & $\qmsr=\cms$ \\
$\hqssxr=\hqssx$ & $\qssxr=\qssx$ & $ $ & $ $ \\
\hline
\multicolumn{2}{c}{$a \neq b$}&\multicolumn{2}{c}{$ $}\\
\hline
$\hqssr=\hqss$ & $\qssr=\qss$ & $ $ & $ $ \\
\hline
\end{tabular}
\label{table:RS2}
\end{table}
Let us notice that, by definition, $Q_{ss}^{aa}=Q_{\sigma \sigma}^{\alpha \alpha}=1$ and thus we do not have to introduce the corresponding conjugate order parameters $\hat{Q}_{ss}^{aa}$ and $\hat{Q}_{\sigma \sigma}^{\alpha \alpha}$.
Similarly, since $Q_{ss\sigma}^{11\alpha}=\mxr$ we do not have to define either $Q_{ss\sigma}^{11\alpha}$ nor $\hat{Q}_{ss\sigma}^{11\alpha}$.
Moreover, we can perform the integration over $\mx$, which fixes $\mx$ to be equal to $2r-1$, using the delta function in Eq.~(\ref{eq:eqIrep}).
Thanks to the RS ansatz, we can easily linearize the quadratic terms in Eq.~(\ref{eq:Phireplicated}) to compute the sums over $\{ \ux \}$, $\{ \ux \}$ and the integrals over $ \underline{\hat{h}} $, $\{ \underline{\hat{\eta}} \}$ and $\{ \underline{\hat{\mu}} \}$. The linearization can be done using the Hubbard-Stratonovich transformation, which introduces the integration variables $\uz$ and $\uxx$, as seen in Eqs.~(\ref{eq:eqdefDZ}) and~ (\ref{eq:eqdefDX}) resulting in three delta functions, which lead to the expressions
\ber
 h_i \: \: \: =\: && z_1 + s_i^{(1)} \Delta_{\hat{h}s}\:, \label{eq:h_iasexpressofzandx} \\
 \eta_i^{(\alpha)}= \: && z_2+x_1+ s_i^{(1)} \Delta_{\hat{\eta}s} + s_i^{(1)} \sigma_i^{(\alpha)} \Delta_{\hat{\eta}s \sigma} \:,  \\
 \mu_i^{(\alpha)}= \: && z_3+x_2+ s_i^{(1)} \Delta_{\hat{\mu}s} + s_i^{(1)} \sigma_i^{(\alpha)} \Delta_{\hat{\mu}s \sigma} \:,
\eer
for the three fields contained in the Heaviside function in Eq.~(\ref{eq:eqIrep}).
According to the value of the spin $s_i^{(1)}$ to be summed over, we define the quantity
\be
\theta^{\pm}_{\uxx, \uz}(\{\Delta\}) = \theta \left[ \left( z_1 \pm \Delta_{\hat{h}s}\right) \left( \frac{}{} z_2+x_1\pm \Delta_{\hat{\eta}s}  \pm  \Delta_{\hat{\eta}s \sigma} + z_3+x_2 \pm \Delta_{\hat{\mu}s}  \pm \Delta_{\hat{\mu}s \sigma} -  z_1 \mp \Delta_{\hat{h}s} \right) \right]\:,
\label{eq:thetapmxzdef}
\ee
where we define
\be
\qquad \{ \Delta \} = \left \{ \begin{array} {cc}
\Delta_{\hat{h}s}=& \hchs- \hqhs \\
\Delta_{\hat{\eta}s}=& \hces - \hqes \\
\Delta_{\hat{\mu}s}=& \hcms - \hqms \\
\end{array}
\label{eq:defDeltaoff}
\right. \:.
\ee
The RS expression of the quantity given in Eq.~(\ref{eq:eqNnm1}) becomes
\be
 \mathds{E}_J \left[  \mathcal{N}_{nm} \right]  = \int \mathcal{D} \mathcal{Q}\hat{\mathcal{Q}} \exp \left[ N \left( - \Psi( \{ Q, \hat{Q} \} ) + \Omega ( \{ Q \}) \right) +\Phi ( \{ \hat{Q} \} )  \right]\:;
\label{eq:NmsRS}
\ee
where the integration measure $\mathcal{D} \mathcal{Q}\hat{\mathcal{Q}}$ contains all the order parameters defined in Tables~\ref{table:RS1} and~\ref{table:RS2}, except $\mx$.
The three functions in the exponent come from the integral representation of the order parameters, the average over disorder and the sum over the spins, respectively.
Their expressions are
\ber
\Psi( \{ Q, \hat{Q} \} )=  && \frac{m(m-1)}{2} \hqee \qee + m \hcee \cee + \frac{m(m-1)}{2} \hqmm \qmm + m \hcmm \cmm + \nonumber  \\
&& + m(m-1) \hqem \qem + m \hcem \cem +\frac{m(m-1)}{2} \hqxx \qxx +   \frac{n(n-1)}{2} \hqss \qss +  \nonumber  \\
&& + m \hche \che + m \hchm \chm+ m \hmx \mx + m \hchsx \chsx +  \hchh \chh \nonumber + \\
&& + m(m-1) \hqesx \qesx + m  \hcesx \cesx + m(m-1) \hqmsx \qmsx + m  \hcmsx \cmsx + \nonumber  \\
&& + \hchs \chs  + (n-1) \hqhs \qhs + m \hces \ces + m(n-1) \hqes \qes + \nonumber \\
&& + m \hcms \cms + m(n-1) \hqms \qms + m(n-1) \hqssx \qssx +\nonumber  \\
&& + \frac{n}{2} \hqss + \frac{m}{2} \hqxx + m \hqssx \mx \:, \label{eq:eqPsirs}
\eer
\ber
\Omega(\{ Q \}) =&&     3\chh+ 6 \chs^2 + 3 m(m-1) \left[  \qee \qxx^2 + 2 \qesx^2 \qxx\right] + 3 m \left[ \cee + 2 \cesx^2\right] + \nonumber \\
&& + 6 m(m-1) \left[ \qmm \qxx + \qmm \mx^2\right] + 6 m \left[ \cmm + \cmm \mx^2 \right] + 6 m(m-1) \qmsx^2 + \nonumber  \\
&& + 6 m \cmsx^2 + 12 m(m-1)\cms \qmsx \mx + 12 m \cms \cmsx \mx + 6 m (m-1) \cms^2 \qxx + \nonumber  \\
&& + 6m \cms^2 + 6 m \che \mx^2 + 12 m \chsx \ces \mx + 12 m \chm \mx + 12 m \chsx \cms +  \nonumber  \\
&& + 12 m \chs \cms \mx + 12 m (m-1) \qee \qxx \mx + 12 m \cem \mx + 12 m \cesx \cmsx \mx  + \nonumber \\
&& + 12 m (m-1) \qesx \qmsx \mx + 12 m (m-1) \ces \qmsx \qxx + 12 m \ces \cmsx + \nonumber \\
&& + \frac{1}{4} \left[\frac{}{} n (n-1) \beta^2 \qss^3 + \beta^2 n  + 2 \beta \left( 3 \chs + 3(n-1) \qhs \qss^2 + 3 m \ces \mx^2 +    \right. \right.   \nonumber \\
&&  \left. \left.  + 3 m (n-1) \qes \qssx^2 + 6 m \cms \mx + 6 m (n-1) \qms \qssx \qss \right) \frac{}{} \right] \:,
\label{eq:eqOmegars}
\eer
\ber
\Phi ( \{ \hat{Q} \}) =  \log && \left[ \int D\uz^{+} e^{z_5 - (z_4 + \hmx)m} \left(1+m g^+(\hmx,\{\Delta\}) \frac{}{}\right) (2\cosh z_5)^{n-1} +\right. \nonumber \\
&&+ \left. \int D\uz^{-} e^{-z_5 - (z_4 + \hmx) m} \left(1+m g^-(\hmx,\{\Delta\}) \frac{}{} \right) (2\cosh z_5)^{n-1}  \right]\:,
\label{eq:eqPhirs}
\eer
and in the last expression, similarly to what has been done in Eq.~(\ref{eq:eqD}), we defined the measure
\ber
&& D\uz^{\pm}=\sqrt{\frac{\det \left( U^{\pm} \right)^{-1} }{ (2\pi)^5}} \prod_{k=1}^5 dz_k \exp \left \{ -\frac{1}{2} \uz^T \left( U^{\pm} \right)^{-1} \uz \right \}\:, \nonumber \\
&& U^{\pm} = \left(
\begin{array}{ccccc}
2 \hchh & \hche & \hchm & \pm \hchsx & \hqhs \\
\hche & \hqee & \hqem & \pm \hqesx & \hqes \\
\hchm & 	\hqem & \hqmm & \pm \hqmsx & \hqms \\
\pm \hchsx & \pm \hqesx  &  \pm \hqmsx & \hqxx & \pm \hqssx \\
\hqhs  & \hqes  & \hqms & \pm \hqssx & \hqss \\
\end{array}
\label{eq:eqdefDZ}
\right)\:,
\eer
and the function
\be
g^{\pm} (\hmx,\{\Delta\}) = \log \left[ 1 + e^{2(z_4 + \hmx)} \int D \uxx \theta^{\pm}_{\uxx, \uz}(\{\Delta\}) \right] \:,
\label{eq:eqdefgpm}
\ee
with the corresponding measure
\be
D\uxx =
\sqrt{\frac{\det  V^{-1} }{ (2\pi)^2}} \prod_{k=1}^2 dx_k \exp \left \{ -\frac{1}{2} \uxx^T V^{-1} \uxx \right \} \:, \qquad
V=\left (
\begin{array}{cc}
\delta_{\hat{\eta} \hat{\eta}} & \Delta_{\hat{\eta} \hat{\mu}} \\
\Delta_{\hat{\eta} \hat{\mu}} & \delta_{\hat{\mu} \hat{\mu}}\\
\end{array}
\right) \:,
\label{eq:eqdefDX}
\ee
where the entries of the matrix $V$ are given by
\be
\begin{array}{cc}
\delta_{\hat{\eta} \hat{\eta}}=&2\hcee - \hqee \\
\delta_{\hat{\mu} \hat{\mu}}=&2\hcmm - \hqmm \\
\Delta_{\hat{\eta} \hat{\mu}}=& \hcem - \hqem \\
\end{array}\:,
\label{eq:deltasmallmeasX}
\ee
and the function $\theta^{\pm}_{\uxx, \uz}(\{\Delta\})$ has been defined in Eq.~(\ref{eq:thetapmxzdef}).
The integral over the inner measure in Eq.~(\ref{eq:eqdefgpm}) can be done analytically, as shown later in Eqs.~(\ref{eq:defipmsscdelta})-(\ref{eq:defipmsscdeltaresult}), and it is equal to
\be
L_{\pm}= \int D \uxx \theta^{\pm}_{\uxx, \uz}(\{\Delta\}) = 1-\frac{1}{2} \erfc \left[ \frac{c_{\pm}}{\sqrt{2Da_{\pm}^2}} \right]\:,
\label{eq:defLthetaintegralssc}
\ee
where $\erfc(x)$ is the complementary error function,
\be
\erfc(x) = 1-\frac{2}{\sqrt{\pi}} \int_{0}^{x} e^{-t^2} dt\:,
\ee
and the parameters appearing in its argument are defined by
\be
D= V_{11}+ V_{22}+2 V_{12}\:,
\label{eq:DcoeffdefsscfunVij}
\ee
\be
a_{\pm} = z_{1} \pm \Delta_{\hat{h}s}\:,
\label{eq:defapmsscfz}
\ee
\be
c_{\pm} =\left[ \frac{}{}( z_2 \pm \Delta_{\hat{\eta}s} \pm \Delta_{\hat{\eta}s \sigma})  + ( z_3 \pm \Delta_{\hat{\mu}s} \pm \Delta_{\hat{\mu}s \sigma}) - (z_1 \pm \Delta_{\hat{h}s}) \right] (z_1 \pm \Delta_{\hat{h}s})\:.
\label{eq:defcpmsscfz}
\ee

Having introduced all these definitions, we can evaluate the integral in Eq.~(\ref{eq:NmsRS}) with the steepest descent method, after the transformation $-i \{ \hat{Q} \} = \{ \hat{Q} \}$, where $\{ Q \}$ is the set of all the order parameters involved.
The saddle point equations obtained by optimizing with respect to the original order parameters read
\ber
\frac{n (n-1)}{2} \hqss = && \frac{3}{4} \beta^2 (n-1) m \qss^2 + \frac{1}{4} \left(\frac{}{} 6 \beta (n-1) \qhs \qss + 12\beta m (n-1) \qms \qssx \frac{}{}\right) \:, \label{eq:eqqssmn}\\
\hqssx = && \frac{3}{2} \left[ 2 \beta \qes \qssx  + \qms \qss \right] \label{eq:speqqss}\:, \\
\hqxx= && 3 \left[ \qee \qxx + \qesx^2 + \qmm  + \cms^2 + 2 \qem \mx  + 2 \ces \qmsx \right] \label{eq:eqsapoihqxx}\:,
\eer
\ber
\hchs = && \frac{1}{4} \left( 6 \beta + 12 \chs + 12 m \cms \mx \right) \:, \label{eq:eqsapoihchs} \\
\hces = && \frac{3}{2}\left( \beta \mx^2 + 2 \chsx \mx + 2(m-1) \qmsx \qxx + 2 \cmsx \frac{}{} \right)\:, \label{eq:eqsapoihces} \\
\hcms = && \frac{3}{2}\left(\frac{}{} \beta \mx + 2 (m-1) \qmsx \mx + 2 \cmsx \mx + 2 (m-1) \cms \qxx + \right. \nonumber \label{eq:eqsapoihcms} \\
&& \left. + 2 \cms + 2 \chsx + 2 \chs \mx \frac{}{}\right)\:,
\eer
\ber
\hqhs = && \frac{3}{2} \beta \qss^2\:,  \label{eq:eqsapoihqhs}  \\
\hqes = && \frac{3}{2} \beta \qssx^2\:,  \label{eq:eqsapoihqes}  \\
\hqms = && 3 \beta \qssx \qss  \label{eq:eqsapoihqms} \:,
\eer
\ber
\hchh = && \frac{3}{4}\:, \\
\hcee = && \frac{3}{4} \:,  \label{eq:eqsapoihcee} \\
\hcmm = && \frac{3}{2}(1+\mx^2) \:, \label{eq:eqsapoihcmm} \\
\hcem = && 3 \mx \:, \label{eq:eqsapoihcem} \\
\hche = && \frac{3}{2} \mx^2\:, \\
\hchm = && 3 \mx \:, \label{eq:hchmsaddlepoint}
\eer
\ber
\hqee = && \frac{3}{2} \qxx^2\:, \label{eq:eqsapoihqee} \\
\hqmm = && 3 (\qxx + \mx^2) \:, \label{eq:eqsapoihqmm} \\
\hqem = &&  3 \qxx \mx \:, \label{eq:eqsapoihqem}
\eer
\ber
\hchsx = && 3 \left[ \ces \mx + \cms \right]\:,  \label{eq:eqsapoihchsx} \\
\hcesx = && 3 \left[ \cesx + \cmsx \mx \right]\:,  \label{eq:eqsapoihcesx}  \\
\hcmsx = && 3 \left[ \cmsx + \cms \mx + \cesx \mx + \ces \right]\:,  \label{eq:eqsapoihcmsx}  \\
\hqesx = && 3 \left[ \qesx \qxx + \qmsx \mx \right]\:, \label{eq:eqsapoihqesx} \\
\hqmsx = && 3 \left[ \qmsx + \cms \mx + \qesx \mx + \ces \qxx \right]\:.  \label{eq:hqmsxsaddlepoint}
\eer
In order to simplify the expressions of next saddle point equations we introduce the notation
\be
D_{ij}=\frac{\partial \Phi}{\partial U^+_{ij}}\:, \qquad F_{ij}=\frac{\partial \Phi}{\partial V_{ij}}\:.
\label{eq:eqdefDFij}
\ee
At zero order in $m$ and $n$, it is easy to see that $D^0_{ij}=F^0_{ij}=0$. Let us also denote by
\be
D^{1,m}_{ij} \equiv  \lim_{n \rightarrow 0} \lim_{m \rightarrow 0}  \partial_m D_{ij} \:,
\ee
\be
D^{1,n}_{ij}  \equiv  \lim_{n \rightarrow 0}  \lim_{m \rightarrow 0}  \partial_n D_{ij} \:,
\ee
\be
F^{1,m}_{ij}  \equiv  \lim_{n \rightarrow 0}  \lim_{m \rightarrow 0}  \partial_m F_{ij} \:
\ee
their $O(m)$ and $O(n)$ contributions. Moreover, it is also useful to introduce
\be
G^{1,m} \equiv  \lim_{n \rightarrow 0} \lim_{m \rightarrow 0} \frac{\partial }{\partial \hat{C}_{\bullet}} \partial_m \Phi\:,
\ee
where we indicate by $\hat{C}_{\bullet}$ any of $\{\hces, \hcms, \hcesx, \hcmsx \}$, and
\be
M^{1,m} \equiv  \lim_{n \rightarrow 0} \lim_{m \rightarrow 0} \frac{\partial }{\partial \hmx} \partial_m \Phi\:.
\ee
The saddle point equations obtained by optimizing with respect to the conjugate order parameters are
\ber
\qss = && 1-2 D_{55}^{1,n} \:, \label{eq:D55spqssssc} \\
\qssx = && \mx -D_{54}^{1,m} \:, \label{eq:spqssxsscf} \\
\qxx = && 1-2 D_{44}^{1,m} \:,
\eer
\ber
\chs = && O(m) \:, \label{eq:speqchssp} \\
\ces = &&  G^{1,m} \:, \label{eq:speqcessp}  \\
\cms = &&  G^{1,m} \:, \label{eq:speqcmssp}
\eer
\ber
\qhs = && O(m) \:, \label{eq:speqqhssp} \\
\qes = && -D_{52}^{1,m} \:, \\
\qms = && -D_{53}^{1,m} \:,
\eer
\ber
\chh = && O(m) \:,  \label{eq:speqchhsp} \\
\cee = && 2 F^{1,m}_{11} \:, \label{eq:speqceesp}  \\
\cmm = && 2 F^{1,m}_{22} \:, \label{eq:speqcmmsp}  \\
\cem = && 2 F^{1,m}_{12} \:, \label{eq:speqcemsp} \\
\che = && -D_{21}^{1,m} \:, \\
\chm = && -D_{31}^{1,m} \:,
\eer
\ber
\qee = && -2 D_{22}^{1,m} \:, \\
\qmm = && -2 D_{33}^{1,m} \:, \\
\qem = && -2 D_{23}^{1,m} \:,
\eer
\ber
\chsx =&& -D_{41}^{1,m} \:, \\
\cesx = &&  G^{1,m} \:, \label{eq:speqcesxsp}  \\
\cmsx= &&  G^{1,m} \:, \label{eq:speqcmsxsp}  \\
\qesx = && -D_{42}^{1,m}\:, \\
\qmsx = && -D_{43}^{1,m} \:,
\eer
\be
\mx =  M^{1,m} \label{eq:spmxsscf} \:.
\ee
The reason why we have one additional equation with respect to Eqs.~(\ref{eq:eqqssmn})-(\ref{eq:hqmsxsaddlepoint}) is that we did not have to optimize over $\mx$ but have to optimize over $\hmx$.
We introduce the term
\be
x_{\pm} \equiv \frac{c_{\pm}}{\sqrt{2Da_{\pm}^2}}\:,
\label{eq:eqdefxpmsqrtD}
\ee
and the functions
\be
f^{\pm}(z_5) = \frac{e^{\pm z_5}}{2 \cosh (z_5)}\:,
\label{eq:fpmsscdefaux}
\ee
to obtain the following expressions for the matrices introduced above
\begin{flalign}
& (i,j) =  \{(2,2),(3,3),(2,3)\}: & \nonumber \\
& D_{ij}^{1,m}= \int D \uz^{+} T^{+}_{ij}(\uz) \left(\frac{}{}-z_4+g^{+}(\hmx, \{ \Delta \})\right) \: f^+(z_5) + &\nonumber \\
& \qquad \quad \int D \uz^{-} T^{-}_{ij} (\uz) \left(\frac{}{}-z_4+g^{-}(\hmx, \{ \Delta \})\right) \:f^-(z_5) +& \nonumber \\
& \qquad \quad \int D \uz^{+} f^+(z_5) \frac{e^{2(z_4+\hmx)}}{e^{g_+}} \frac{e^{-x_+^2}}{\sqrt{\pi}} \frac{x_+}{2D}\alpha +
 \int D \uz^{-} f^-(z_5) \frac{e^{2(z_4+\hmx)}}{e^{g_-}} \frac{e^{-x_{-}^2}}{\sqrt{\pi}} \frac{x_-}{2D}\alpha  \:, \nonumber & \\
& \qquad \quad \textrm{where} \quad \alpha=
\left \{ \begin{array}{l l}
1 & (i,j) = \{(2,2),(3,3)\} \\
2 & (i,j) = \{(2,3)\} \\
\end{array} \right.\:, &
\end{flalign}
\begin{flalign}
&(i,j) =  \{ (4,2),(4,3),(5,2),(5,3)\} :  & \nonumber  \\
& D_{ij}^{1,m} =  \int D \uz^+ T^+_{ij}(\uz) \left(\frac{}{}-z_4+g^{+}(\hmx, \{ \Delta \})\right) f^+(z_5) + & \nonumber \\
& \qquad \quad \alpha \int D \uz^- T^-_{ij}(\uz) \left(\frac{}{}-z_4+g^{-}(\hmx, \{ \Delta \})\right) f^-(z_5) + & \nonumber \\
& \qquad \quad- \int D \uz^{+} f^+(z_5)   \frac{e^{2(z_4+\hmx)}}{e^{g_+}} \frac{e^{-x_+^2}}{\sqrt{\pi}} \frac{\sgn(a_+)}{\sqrt{2D}} +
 \int D \uz^{-} f^-(z_5)   \frac{e^{2(z_4+\hmx)}}{e^{g_-}} \frac{e^{-x_-^2}}{\sqrt{\pi}} \frac{\sgn(a_-)}{	\sqrt{2D}} \:, & \nonumber \\
&  \qquad \quad \textrm{where}\quad \alpha=
\left \{ \begin{array}{cc}
1& (i,j)=\{(5,2),(5,3)\} \\
-1& (i,j)=\{(4,2),(4,3)\} \\
\end{array} \right.\:, 	&
\end{flalign}
\begin{flalign}
& (i,j) =  \{ (2,1),(3,1),(4,4)\} :  & \nonumber  \\
& D_{ij}^{1,m} =  \int D \uz^+ T^+_{ij}(\uz) \left(\frac{}{}-z_4+g^{+}(\hmx, \{ \Delta \})\right) f^+(z_5) + & \nonumber \\
& \qquad \quad \int D \uz^- T^-_{ij}(\uz) \left(\frac{}{}-z_4+g^{-}(\hmx, \{ \Delta \})\right) f^-(z_5)\:, &
\end{flalign}
\begin{flalign}
& (i,j)=\{(4,1),(5,4)\}: \nonumber & \\
& D_{ij}^{1,m} = \int D \uz^+ T^+_{ij}(\uz) \left(\frac{}{}-z_4+g^{+}(\hmx, \{ \Delta \})\right) f^+(z_5) + &\nonumber \\
& \qquad \quad - \int D \uz^- T^-_{ij}(\uz) \left(\frac{}{}-z_4+g^{-}(\hmx, \{ \Delta \})\right) f^-(z_5)\: , \label{eq:D4154sscspeq}&
\end{flalign}
where the terms $T_{ij}^{\pm}(\uz)$ are given by
\be
T_{ij}^{\pm}(\uz)= \left \{ \begin{array}{cc}
\sum_{lp} [U^{\pm}]^{-1}_{li} [U^{\pm}]^{-1}_{pj} z_l z_p - [U^{\pm}]^{-1}_{ij} & i \neq j \\
 & \\
\frac{1}{2} \sum_{lp} [U^{\pm}]^{-1}_{li} [U^{\pm}]^{-1}_{pi} z_l z_p - \frac{1}{2} [U^{\pm}]^{-1}_{ii} & i=j \\
\end{array}
\right.  \:,  \qquad \uz=\{z_1, z_2, z_3, z_4, z_5\} \:,\label{eq:defTijz} \\
\ee
and the matrix $U^{\pm}$ has been defined in Eq.~(\ref{eq:eqdefDZ}). Above we have the expressions of $12$ terms but the matrix $D_{ij}$, defined in Eq.~(\ref{eq:eqdefDFij}) should contain $15$ independent entries. One of the missing terms is $D_{55}$ that will be described shortly, while the other two are $D_{11}$ and $D_{51}$ which account for the $O(m)$ contributions in Eqs.~(\ref{eq:speqqhssp}) and~(\ref{eq:speqchhsp}) to be discussed later. The term $D_{55}$ appears in Eq.~(\ref{eq:D55spqssssc}) but we are interested in its $O(n)$, rather than its $O(m)$, contribution:
\be
D_{55}^{1,n}= \frac{1}{2}-\frac{1}{2} \int D t \tanh^2 \left( t \sqrt{\hqss} \right)\:.
\ee
We notice that with this expression, Eqs.~(\ref{eq:eqqssmn}) and~(\ref{eq:D55spqssssc}) correctly describe the reference system in the RS phase only under the hypothesis $\qhs=0$, $\qms \qssx=0$.
In fact, the reference system should not be affected by the order parameters related to the computation of the SSC entropy, following the arguments made in the computation of the Franz-Parisi potential.
Our approach to deal with this problem is to assume $\qhs=0$, $\qssx=0$ and check that these conditions lead to a  self-consistent result.
Before addressing this problem, we give the explicit expressions for the other matrix terms appearing in the equations above.
The three terms $F_{ij}^{1.m}$ appearing in Eqs.~(\ref{eq:speqceesp})-(\ref{eq:speqcemsp}) are given by
\ber
&& F^{1,m}_{ij}= \int D \uz^{+} \: f^+(z_5) \frac{\partial g^{+}}{\partial V_{ij}} +  \int D \uz^{-} \: f^-(z_5) \frac{\partial g^{-}}{\partial V_{ij}} \:,  \\
&& \frac{\partial g^{\pm}}{\partial V_{ij}} = \left \{ e^{g^{\pm}} \right \}^{-1} \left [ e^{2(z_4 + \hmx)} \int D \uxx \: t_{ij} (\uxx) \theta^{\pm}_{\uxx, \uz}(\{\Delta \})  \right ]\:;  \\
\eer
where
\ber
t_{ij}(\uxx)= \left \{ \begin{array}{cc}
\sum_{lp} [V]^{-1}_{li} [V]^{-1}_{pj} x_l x_p - [V]^{-1}_{ij} & i \neq j \\
 & \\
\frac{1}{2} \sum_{lp} [V]^{-1}_{li} [V]^{-1}_{pi} x_l x_p - \frac{1}{2} [V]^{-1}_{ii} & i=j \\
\end{array}
\right.  \:;  \qquad \uxx=\{x_1, x_2\}\:.  \\
\nonumber
\eer
We introduced the term $G^{1,m}$ because differentiating $\Phi(\{\hat{Q}\})$ with respect to $\hces$, $\hcms$, $\hcesx$, $\hcmsx$ results in the same expression, as can be seen from Eq.~(\ref{eq:thetapmxzdef}). This quantity is given by
\ber
G^{1,m}= && \int D \uz^{+} f^+(z_5) i_+(\uz) \frac{e^{2(\hmx + z_4)}}{e^{g_+}} (z_1 + \Delta_{\hat{h}s} )+  \nonumber \\
&& -  \int D \uz^{-} f^-(z_5) i_-(\uz) \frac{e^{2(\hmx + z_4)}}{e^{g_-}} (z_1 - \Delta_{\hat{h}s} ) \:,   \\
\eer
and it appears in Eqs.~(\ref{eq:speqcessp}), (\ref{eq:speqcmssp}), (\ref{eq:speqcesxsp}) and (\ref{eq:speqcmsxsp}). The integrals $i_{\pm}(\uz)$, introduced above, can be defined by replacing the Heaviside function in Eq.~(\ref{eq:defLthetaintegralssc}) by a Dirac delta,
\be
i_{\pm}(\uz)=\int D \uxx \delta^{\pm}_{\uz, \uxx} (\{\Delta\})\:,
\label{eq:defipmsscdelta}
\ee
and are given by
\ber
i_{\pm}(\uz)=\frac{1}{\sqrt{2\pi D} |a_{\pm}|} \exp \left[ -\frac{1}{2} D^{-1} \left(\frac{c_{\pm}}{a_{\pm}} \right)^2 \right]~.
\label{eq:defipmsscdeltaresult}
\eer
The parameters $a_{\pm}$, $c_{\pm}$ have been defined in Eqs.~(\ref{eq:defapmsscfz})-(\ref{eq:defcpmsscfz}). Finally, the term $M^{1,m}$, appearing in Eq.~(\ref{eq:spmxsscf}), is given by
\be
\label{eq:eqdefM}
M^{1,m}=-1 + \int D \uz^{+} \: f^+(z_5) \frac{\partial g^{+}}{\partial \hmx} +  \int D \uz^{-} \: f^-(z_5) \frac{\partial g^{-}}{\partial  \hmx} \:,
\ee
where the internal derivative will be discussed later.

At this point let us notice that we have $24$ saddle point equations arising from the optimization with respect to the conjugate order parameters, see Eqs.~(\ref{eq:D55spqssssc})-(\ref{eq:spmxsscf}), but above we provided only $13$ expressions for $D_{ij}$, $3$ expressions for $F_{ij}$, the expression for $G^{1,m}$ appearing in $4$ equations, and the expression for $M^{1,m}$, describing in total $21$ of the saddle point equations.
In other words we still need to describe the three equations~(\ref{eq:speqchssp}),~(\ref{eq:speqqhssp}) and (\ref{eq:speqchhsp}). As long as $T>T_K$, we can safely set $\hqss=\qss=0$; moreover, since we adopted the ansatz $\qhs=\qssx=0$ using insight from the Franz-Parisi derivation, Eqs.~(\ref{eq:speqqss}), (\ref{eq:eqsapoihqhs}), (\ref{eq:eqsapoihqes}) and (\ref{eq:eqsapoihqms}) lead to $\hqssx=0$ and $\hqhs=\hqes=\hqms=0$.
These simplifications allows one to write the matrix $U^{\pm}$, defined in Eq.~(\ref{eq:eqdefDZ}), in the following way
\be
  \renewcommand{\arraystretch}{0.6}
  U^{\pm}=\lim_{\hqss\rightarrow 0}\left(
  \begin{array}{ c c c c c }
    \multicolumn{1}{c}{} \quad & \quad & \quad & \quad & 0 \\
    \multicolumn{1}{c}{} \quad & \quad & \quad & \quad & 0 \\
    \multicolumn{1}{c}{} \quad & \quad & \quad & \quad & 0 \\
    \multicolumn{4}{c}{\raisebox{1\normalbaselineskip}[0pt][0pt]{$U^{\pm}_4$}} & 0 \\
    0 & 0 & 0 & 0 & \hqss \\
     \end{array}
     \label{eq:eqdefDZRS}
  \right)
\:, \qquad
U_4^{\pm}=
\left(
\begin{array}{cccc}
2 \hchh & \hche & \hchm & \pm \hchsx  \\
\hche & \hqee & \hqem & \pm \hqesx  \\
\hchm & 	\hqem & \hqmm & \pm \hqmsx  \\
\pm \hchsx & \pm \hqesx  &  \pm \hqmsx & \hqxx  \\
\end{array}
\right)\:.
\ee
While checking the self-consistency of $\qss=\hqss=0$ is very easy, it is slightly more involved to prove that $\qssx=\qhs=0$ are self-consistent solutions as well.
Let us first consider the condition $\qssx=0$ and notice that all we need to show is the equality $D_{54}^{1,m}=M^{1,m}$, as follows from Eqs.~(\ref{eq:spqssxsscf}) and~(\ref{eq:spmxsscf}).
The expression $M^{1,m}$, given in Eq.~(\ref{eq:eqdefM}), contains an inner derivative that we did not discuss so far. Given the symmetric role of $\hmx$ and $z_4$ in $g^{\pm}$ in Eq.~(\ref{eq:eqdefgpm}), we have
\be
\frac{\partial g^{\pm} (\hmx,\{\Delta\})}{\partial \hmx} = \frac{\partial g^{\pm} (\hmx,\{\Delta\})}{\partial z_4} \:.
\ee
Integrating by parts and differentiating the Gaussian measure, we get
\ber
M^{1,m}=-1 + && \int D \uz^+ \left[ \sum_{k} [U^+]^{-1}_{4k} z_k \right] f^+(z_5) g^+(\hmx,\{\Delta\}) + \\
&& \int D \uz^- \left[ \sum_{k} [U^-]^{-1}_{4k} z_k \right] f^-(z_5) g^-(\hmx,\{\Delta\}) \:.
\label{eq:M1maniprsssc}
\eer
Using Eq.~(\ref{eq:eqdefDZRS}) we can isolate the measure on $z_5$ in $D\uz^{\pm}$,
\be
D\uz^{\pm} \equiv D\uz_4^{\pm} Dz_5 \:, \qquad Dz_5 = \sqrt{\frac{\hqss^{-1}}{2\pi}} \exp{\left[ -\frac{1}{2} z_5^2 \hqss^{-1} \right]} d z_5
\label{eq:decoupleDz4dz5}
\ee
and, given the definition in Eq.~(\ref{eq:fpmsscdefaux}), it is easy to prove that
\be
\lim_{\hqss \rightarrow 0 } \int D z_5 f^{\pm}(z_5) = \frac{1}{2}\:.
\ee
Using this equality in Eq.~(\ref{eq:M1maniprsssc}) we obtain
\ber
M^{1,m}=-1 + && \frac{1}{2} \int D \uz_4^+ \left[ \sum_{k} [U^+]^{-1}_{4k} z_k \right]  g^+(\hmx,\{\Delta\}) +  \nonumber  \\
&&  \frac{1}{2} \int D \uz_4^- \left[ \sum_{k} [U^-]^{-1}_{4k} z_k \right] g^-(\hmx,\{\Delta\})
\label{eq:M1mfinalressscrs}
\eer
and now we need to show that $D_{54}^{1,m}$ has the very same expression under the considered ansatz. This can be done by noticing that the inverse of the matrix $U^{\pm}$ is block diagonal as well
\be
  \renewcommand{\arraystretch}{0.6}
  [U^{\pm}]^{-1}=\lim_{\hqss\rightarrow 0}\left(
  \begin{array}{ c c c c c }
    \multicolumn{1}{c}{} \quad & \quad & \quad & \quad & 0 \\
    \multicolumn{1}{c}{} \quad & \quad & \quad & \quad & 0 \\
    \multicolumn{1}{c}{} \quad & \quad & \quad & \quad & 0 \\
    \multicolumn{4}{c}{\raisebox{1\normalbaselineskip}[0pt][0pt]{$[U^{\pm}_4]^{-1}$}} & 0 \\
    0 & 0 & 0 & 0 & \hqss^{-1} \\
     \end{array}
     \label{eq:eqdefDZRSinv}
  \right)
  \ee
and thus, since $[U^{\pm}]^{-1}_{54}=0$, Eq.~(\ref{eq:defTijz}) leads to
\be
T_{54}^{\pm}(\uz) =  \lim_{\hqss \rightarrow 0} \sum_p [U^{\pm}]^{-1}_{4p} z_p \hqss^{-1} z_5 \:.
\ee
The term $D^{1,m}_{54}$ is given by Eq.~(\ref{eq:D4154sscspeq}) and, decoupling $Dz_5$ and $D\uz_4^{\pm}$ as in Eq.~ (\ref{eq:decoupleDz4dz5}), we can write its two contributions as
\be
 \int D\uz^{\pm} T_{54}^{\pm} (\uz) g^{\pm} (\hmx,\{\Delta\}) f^{\pm}(z_5) =  \int D \uz_4^{\pm} D z_5  \sum_p [U^{\pm}]^{-1}_{4p} z_p \hqss^{-1} z_5 g^{\pm} (\hmx,\{\Delta\})f^{\pm}(z_5) \:,
 \ee
 \be
 \int D \uz^{\pm} T_{54}^{\pm} (\uz) z_4  f^{\pm}(z_5) =  \int D \uz_4^{\pm} D z_5  \sum_p [U^{\pm}]^{-1}_{4p} z_p \hqss^{-1} z_5 z_4 f^{\pm}(z_5)\:.
\ee
We can integrate over $z_5$ in both the right hand sides of the equations
\be
\lim_{\hqss \rightarrow 0 } \int D z_5 \hqss^{-1} z_5 f^{\pm}(z_5) = \pm \frac{1}{2}\:,
\ee
because $g^{\pm} (\hmx,\{\Delta\})$ does not depend on $z_5$ we obtain
\ber
D^{1,m}_{54} = && - \frac{1}{2} \int D \uz^+_{4} \sum_p [U^+]^{-1}_{4p} z_p z_4 - \frac{1}{2} \int D \uz^-_{4} \sum_p [U^-]^{-1}_{4p} z_p z_4 + \\
&& + \frac{1}{2} \int  D \uz^+_{4} \sum_p [U^+]^{-1}_{4p} z_p  g^{+} (\hmx,\{\Delta\}) + \frac{1}{2} \int  D \uz^-_{4} \sum_p [U^-]^{-1}_{4p} z_p  g^{-} (\hmx,\{\Delta\})~. \nonumber
\eer
Finally, integrating over the four dimensional measure the product $z_p z_4$, we obtain
\ber
D^{1,m}_{54}=-1 + && \frac{1}{2} \int D \uz_4^+ \left[ \sum_{k} [U^+]^{-1}_{4k} z_k \right]  g^+(\hmx,\{\Delta\}) +  \nonumber  \\
&&  \frac{1}{2} \int D \uz_4^- \left[ \sum_{k} [U^-]^{-1}_{4k} z_k \right] g^-(\hmx,\{\Delta\})
\eer
and so we recognize that $D^{1,m}_{54} = M^{1,m}$, see Eq.~(\ref{eq:M1mfinalressscrs}). This strategy can be used also to prove that the zero order terms in the equations leading to $\chs$, $\qhs$ and $\chh$ is zero, i.e. to validate Eqs.~(\ref{eq:speqchssp}), (\ref{eq:speqqhssp}) and (\ref{eq:speqchhsp}), where the second is needed to check the self-consistency of the ansatz considered.

These manipulations lead to a simplified system of equations, that contains \emph{only} 9 equations, compared to the $23+24=47$ equations we started with, as describe below.
In fact, plugging Eqs.~(\ref{eq:eqPsirs}), (\ref{eq:eqOmegars}) and (\ref{eq:eqPhirs}) in Eq.~(\ref{eq:NmsRS}) and using this equation in Eq.~(\ref{eq:eqdefEntr}), we end up with an expression of $S_{\beta}(r)$ that can be written in terms of the original order parameters (i.e. without the conjugate ones) thanks to Eqs.~(\ref{eq:eqqssmn})-(\ref{eq:hqmsxsaddlepoint}). In the present ansatz this expression reduces to
\ber
S_{\beta}(r)  =  && \frac{3}{2} \qmsx^2 -\frac{3}{2} \qesx^2 + 6 \qmsx \qxx G^{1,m} + 3 [G^{1,m}]^2 (\qxx-1) + \frac{3}{2} \qxx \qmm + \nonumber \\
&& +\frac{3}{2} \qee \qxx^2 + 4 \qem \qxx \mx -\frac{3}{2} \qee \qxx -\frac{3}{2} \qmm -3 \qem \mx -3 G^{1,m} \qmsx  + \nonumber \\
&&  +3 \qesx^2 \qxx + 3 G^{1,m} \mx \qmsx + 3 \qesx \qmsx \mx - 6 [G^{1,m}]^2 (1+\mx) +  \nonumber \\
&&- 3 [G^{1,m}] \mx \chsx -3 G^{1,m} \chsx - \hmx \mx + \partial_m \Phi\:,
\label{eq:finalSbetarsscpsp}
\eer
where the last term is given by
\be
\partial_m \Phi = -\hmx + \frac{1}{2} \int D z_4^+ g^{+} (\hmx,\{\Delta\}) + \frac{1}{2} \int D z_4^- g^{-} (\hmx,\{\Delta\})\:.
\ee
As a reminder, $\mx$ is fixed to be $2r-1$, relating to the size of the SSC. The four dimensional measure $D \uz_4^{\pm}$ can also be partially expressed in terms of the original order parameters and reads
\be
\label{eq:simplyu4rsanssscfin}
U_4^{\pm}=
\left(
\begin{array}{cccc}
\frac{3}{2} & \frac{3}{2} \mx^2 & 3 \mx  & \pm \hchsx  \\
\frac{3}{2} \mx^2 & \frac{3}{2} \qxx^2 & 3 \qxx \mx & \pm \hqesx  \\
3 \mx & 	3 \qxx \mx & 3 (\qxx + \mx^2) & \pm \hqmsx  \\
\pm \hchsx & \pm \hqesx  &  \pm \hqmsx & \hqxx  \\
\end{array}
\right)\:.
\ee
The $9$ order parameters that need to be determined in order to compute the entropy of SSC at size $r$ are
\be
\mx, \quad \qxx, \quad G^{1,m},\quad  \chsx, \quad \qesx,\quad  \qmsx, \quad \qee, \quad \qmm, \quad \qem\:,
\label{eq:listOPfinalcompSSCent}
\ee
as can be seen in Eq.~(\ref{eq:finalSbetarsscpsp}). In what follows we  show how to compute these order parameters self-consistently. Notice that once this list is known, using Eqs.~(\ref{eq:eqsapoihqxx}), (\ref{eq:eqsapoihchsx}), (\ref{eq:eqsapoihqesx}) and (\ref{eq:hqmsxsaddlepoint}), we obtain
\ber
\hqxx = && 3 (\qee \qxx + \qesx^2 + \qmm + [G^{1,m}]^2) + 6 (\qem \mx + G^{1,m} \qmsx) \:, \\
\hchsx = && 3 G^{1,m} (1+\mx) \:, \\
\hqesx = && 3 (\qesx \qxx + \qmsx \mx) \:, \\
\hqmsx = && 3 ( \qmsx + \qesx \mx + G^{1,m} (\mx + \qxx) )  \:,
\eer
and so the matrix $U_4^{\pm}$ in Eq.~(\ref{eq:simplyu4rsanssscfin}) is fully specified.
Moreover, $\mx$ and $\qxx$ completely specify $V$ as well. In fact, using Eqs.~(\ref{eq:eqdefDX}) and~(\ref{eq:deltasmallmeasX}), thanks to Eqs.~(\ref{eq:eqsapoihcee}), (\ref{eq:eqsapoihcmm}), (\ref{eq:eqsapoihcem}), (\ref{eq:eqsapoihqee}), (\ref{eq:eqsapoihqmm}) and (\ref{eq:eqsapoihqem}), we obtain
\be
V=\left (
\begin{array}{cc}
\frac{3}{2}(1-\qxx^2) & 3 \mx (1-\qxx) \\
3 \mx (1-\qxx) & 3(1-\qxx) \\
\end{array}
\right)
\ee
so that $D$ is well defined as well from Eq.~(\ref{eq:DcoeffdefsscfunVij}).
Using Eq.~(\ref{eq:eqsapoihchs}), (\ref{eq:eqsapoihces}), (\ref{eq:eqsapoihcms}), (\ref{eq:eqsapoihcesx}) and (\ref{eq:eqsapoihcmsx}), the $9$ parameters above determine also the following order parameters
\ber
\hchs = && \frac{3}{2}  \beta  \\
\hces = && \frac{3}{2} \beta \mx^2 + 3 \chsx \mx - 3 \qmsx \qxx + 3 G^{1,m} \:, \\
\hcms = && \frac{3}{2} \beta \mx - 3 \qmsx \mx + 3 G^{1,m} \mx - 3 G^{1,m} \qxx +  3 G^{1,m} + \chsx \:, \\
\hcesx  = && 3 G^{1,m} (1+\mx) \:,    \\
\hcmsx = && 6 G^{1,m} (1+\mx)   \:.
\eer
Remembering that $\hqhs=\hqes=\hqms=0$, we can compute $c^{\pm}$ using Eq.~(\ref{eq:defcpmsscfz}), and $x_{\pm}$ using Eq.~ (\ref{eq:eqdefxpmsqrtD}). The term
$L_{\pm}$ can be defined as well and so we have everything we need to compute $g^{\pm}(\hmx,\{\Delta\})$.
All we need to do at this point is to solve self-consistently the equations for the $9$ order parameters specified above.
For the sake of readability, we report all these interlinked equations below:
\ber
\mx = && -1 + \frac{1}{2} \int D \uz_4^+ \left[ \sum_{k} [U^+]^{-1}_{4k} z_k \right]  g^+(\hmx,\{\Delta\}) + \nonumber \\
&&  \qquad \; \; \frac{1}{2} \int D \uz_4^- \left[ \sum_{k} [U^-]^{-1}_{4k} z_k \right] g^-(\hmx,\{\Delta\}) \:, \label{eq:firstfinaleqsscit}\\
\qxx= &&  1 -  \frac{1}{2} \int D \uz_4^+ T^{4,+}_{4,4}(\uz) g^+(\hmx,\{\Delta\})  - \int D \uz_4^- T^{4,-}_{4,4}(\uz) g^-(\hmx,\{\Delta\}) \:, \\
G^{1,m} = &&  \frac{1}{2}\int D \uz_4^+ \frac{e^{2(\hmx + z_4)}}{e^{g_+}} i_+(\uz) (z_1 + \hchs) -  \frac{1}{2}\int D \uz_4^- \frac{e^{2(\hmx + z_4)}}{e^{g_-}} i_-(\uz) (z_1 - \hchs)\:,
\eer

\ber
\chsx = && \frac{1}{2} \int D \uz_4^+ T^{4,+}_{4,1}(\uz)  g^+(\hmx,\{\Delta\})  - \frac{1}{2} \int D \uz_4^- T^{4,-}_{4,1}(\uz)  g^-(\hmx,\{\Delta\})  \:, \\
\qesx = && - \frac{1}{2} \int D \uz_4^+ T^{4,+}_{4,2}(\uz)  g^+(\hmx,\{\Delta\})  + \frac{1}{2} \int D \uz_4^- T^{4,-}_{4,2}(\uz)  g^-(\hmx,\{\Delta\}) \nonumber  \\
&& + \frac{1}{2} \int D \uz_4^{+}  \frac{e^{2(z_4+\hmx)}}{e^{g_+}} \frac{e^{-x_+^2}}{\sqrt{\pi}} \frac{\sgn(a_+)}{\sqrt{2D}} -
\frac{1}{2} \int D \uz^{-}_4    \frac{e^{2(z_4+\hmx)}}{e^{g_-}} \frac{e^{-x_-^2}}{\sqrt{\pi}} \frac{\sgn(a_-)}{	 \sqrt{2D}} \:, \\
\qmsx = &&  - \frac{1}{2} \int D \uz_4^+ T^{4,+}_{4,3}(\uz)  g^+(\hmx,\{\Delta\})  + \frac{1}{2} \int D \uz_4^- T^{4,-}_{4,3}(\uz)  g^-(\hmx,\{\Delta\}) \nonumber  \\
&& + \frac{1}{2} \int D \uz_4^{+}  \frac{e^{2(z_4+\hmx)}}{e^{g_+}} \frac{e^{-x_+^2}}{\sqrt{\pi}} \frac{\sgn(a_+)}{\sqrt{2D}} -
\frac{1}{2} \int D \uz^{-}_4    \frac{e^{2(z_4+\hmx)}}{e^{g_-}} \frac{e^{-x_-^2}}{\sqrt{\pi}} \frac{\sgn(a_-)}{	 \sqrt{2D}} \:,
\eer

\ber
\qee =  && - \int D \uz_4^{+} T^{4,+}_{22}(\uz) g^{+}(\hmx, \{ \Delta \}) - \int D \uz_4^{-} T^{4,-}_{22}(\uz) g^{-}(\hmx, \{ \Delta \}) + \nonumber \\
&&  - \int D \uz_4^{+}  \frac{e^{2(z_4+\hmx)}}{e^{g_+}} \frac{e^{-x_+^2}}{\sqrt{\pi}} \frac{x_+}{2D} - \int D \uz_4^{-}  \frac{e^{2(z_4+\hmx)}}{e^{g_-}} \frac{e^{-x_+^2}}{\sqrt{\pi}} \frac{x_-}{2D}\:,  \\
\qmm =  && - \int D \uz_4^{+} T^{4,+}_{33}(\uz) g^{+}(\hmx, \{ \Delta \}) - \int D \uz_4^{-} T^{4,-}_{33}(\uz) g^{-}(\hmx, \{ \Delta \}) + \nonumber \\
&&  - \int D \uz_4^{+}  \frac{e^{2(z_4+\hmx)}}{e^{g_+}} \frac{e^{-x_+^2}}{\sqrt{\pi}} \frac{x_+}{2D} - \int D \uz_4^{-}  \frac{e^{2(z_4+\hmx)}}{e^{g_-}} \frac{e^{-x_+^2}}{\sqrt{\pi}} \frac{x_-}{2D}\:,  \\
\qem =  && - \frac{1}{2} \int D \uz_4^{+} T^{4,+}_{23}(\uz) g^{+}(\hmx, \{ \Delta \}) - \frac{1}{2} \int D \uz_4^{-} T^{4,-}_{23}(\uz) g^{-}(\hmx, \{ \Delta \}) + \nonumber \\
&&  - \int D \uz_4^{+}  \frac{e^{2(z_4+\hmx)}}{e^{g_+}} \frac{e^{-x_+^2}}{\sqrt{\pi}} \frac{x_+}{2D} - \int D \uz_4^{-}  \frac{e^{2(z_4+\hmx)}}{e^{g_-}} \frac{e^{-x_+^2}}{\sqrt{\pi}} \frac{x_-}{2D}\:, \label{eq:lastfinaleqsscit}
\eer
where, consistently with the notation adopted in Eq.~(\ref{eq:defTijz}), we define
\be
T_{ij}^{4,\pm}(\uz)= \left \{ \begin{array}{cc}
\sum_{lp} [U_4^{\pm}]^{-1}_{li} [U_4^{\pm}]^{-1}_{pj} z_l z_p - [U_4^{\pm}]^{-1}_{ij} & i \neq j \\
 & \\
\frac{1}{2} \sum_{lp} [U_4^{\pm}]^{-1}_{li} [U_4^{\pm}]^{-1}_{pi} z_l z_p - \frac{1}{2} [U_4^{\pm}]^{-1}_{ii} & i=j \\
\end{array}
\right.  \:,  \qquad \uz=\{z_1, z_2, z_3, z_4\} \:.\label{eq:defT4ijz} \\
\ee
This is a complicated system of equations but can solve numerically by iteration.
For each given value of $\hmx$ and $\beta$, we start with some trial values of the order parameters listed in Eq.~(\ref{eq:listOPfinalcompSSCent}).
We compute all the quantities discussed above in order to perform the integrals in equations (\ref{eq:firstfinaleqsscit})-(\ref{eq:lastfinaleqsscit}) and end up with a new list of order parameters.
Integrals over the Gaussian measures are computed using Monte-Carlo sampling. The number of random $4$-dimensional vectors that we need to sample to get reliable results is $O(10^8)$. We iterate this process until convergence by computing the difference between the old and new order parameters starting from different initial conditions.
The first equation provides the value $r$ which corresponds to the initial choice of $\hmx$ and $\beta$.
The corresponding entropy is then provided from Eq.~(\ref{eq:finalSbetarsscpsp}).
Usually, $O(200)$ iteration steps are required to get a reliable solution. We never find multiple solutions starting from different initial conditions but the quality of the solution is found to depend on $r$. In particular, when $r$ is very close to $1$, the error on the final result is larger because it becomes increasingly more difficult to invert the matrices $U$ and $V$. As Fig.~(\ref{fig:entropy}) in the main text shows, this is not a practical problems, since as long as solutions can be found the profile of the entropy $S_\beta(r)$ is very smooth.

Finally, we would like to discuss how the distribution of local fields in SSC can be computed, recycling all the effort made up to now.
To this aim, let us introduce a variant of the original cluster entropy
\be
N S_{\beta}^{\gamma}(r) = \mathds{E}_J \mathds{E}_{\us} \log \sum_{\ux} \mathds{I}^{\gamma}_{\ux}(\us) \: \delta \left(r N - \sum_{i=1}^N \frac{1+\sigma_i}{2}\right)
\ee
where $\mathds{I}^{\gamma}_{\ux}(\us)$, given Eq.~(\ref{eq:eqInd}), is defined by
\be
\mathds{I}^{\gamma}_{\ux}(\us) = \prod_{i=1}^{N} \left \{ \frac{1-\sigma_i}{2} + \frac{1+\sigma_i}{2} \: \theta \left[ u_i^2 - (v_i + w_i)^2 -\epsilon \right] e^{\gamma \delta(h_i-\lambda) }\right \}\:.
\label{eq:eqIndtilted}
\ee
This quantity has two important properties. The trivial one is that $\lim_{\gamma \rightarrow 0}S_{\beta}^{\gamma}(r) = S_{\beta}(r)$.
The less trivial one is that the derivative of this function leads to Eq.~(\ref{Prh}). In fact, let us compute
\be
N \lim_{\gamma \rightarrow 0}  \partial_{\gamma} S_{\beta}^{\gamma}(r) =  \mathds{E}_J \mathds{E}_{\us} \left[ \sum_{\ux} \mathds{I}^{r}_{\ux}(\us) \right]^{-1}
\sum_{\ux,i} \mathds{I}^{(i),r}_{\ux}(\us) \left( \frac{1+\sigma_i}{2} \right) \theta \left[ u_i^2 - (v_i + w_i)^2 -\epsilon \right] \delta(h_i-\lambda)
\label{NderSbetagamma}
\ee
where $\mathds{I}^{r}_{\ux}(\us)$ has been defined in Eq.~(\ref{Prh2}) and $\mathds{I}^{(i),r}_{\ux}(\us)$ is given by
\be
\mathds{I}^{(i),r}_{\ux}(\us)  = \mathds{I}^{(i)}_{\ux}(\us) \delta \left(rN -\sum_{i=1}^{N} \frac{1+\sigma_i}{2}\right) \:,
\label{eq:indiexcspini}
\ee
where
\be
\mathds{I}^{(i)}_{\ux}(\us) = \prod_{j \neq i} \left [ \frac{1-\sigma_j}{2} + \frac{1+\sigma_j}{2} \theta \left[ u_j^2 - (v_j + w_j)^2 -\epsilon \right] \right] \:.
\ee
To recognize that the right hand side of Eq.~(\ref{NderSbetagamma}) is strictly related to Eq.~(\ref{Prh}), we only need to show that
\be
\mathds{I}^{(i),r}_{\ux}(\us) \left( \frac{1+\sigma_i}{2} \right) \theta \left[ u_i^2 - (v_i + w_i)^2 -\epsilon \right] = \mathds{I}^{r}_{\ux}(\us) \left( \frac{1+\sigma_i}{2} \right) \:.
\label{eq:eqdertoprodlocf}
\ee
This equality can be verified as follows.
Firstly, notice that if $\sigma_1=-1$, Eq.~(\ref{eq:eqdertoprodlocf}) is trivially satisfied.
Secondly, the indicator function in $ \mathds{I}^{r}_{\ux}(\us)$ given in Eq.~(\ref{eq:eqInd}), is one if and only if the cluster specified by $\ux$ (i.e. those spins where $\sigma_i=1$) is self-sustained; the same condition holds for $\mathds{I}^{(i)}_{\ux}(\us)$ from the definition in Eq.~(\ref{eq:indiexcspini}), so that it is one if and only if the cluster specified by $\ux \backslash \sigma_i$  is self-sustained.
Thus, even when $\sigma_1=1$, we see that $\mathds{I}^{(i)}_{\ux}(\us)$ on the left hand side lacks a condition on the spin $s_i$ that is explicitly enforced by the Heaviside function.
Thus, Eq.~(\ref{NderSbetagamma}) leads to
\be
r^{-1} \lim_{\gamma \rightarrow 0}  \partial_{\gamma} S_{\beta}^{\gamma}(r)  = P_r(h=\lambda)\:.
\ee
Numerically, we used a small value of $\gamma=0.1$ and solved the two sets of saddle point equations with $\gamma=0$ and $\gamma \neq 0$. In the second case, all we need to do is replace Eq.~(\ref{eq:eqInd}) by Eq.~(\ref{eq:eqIndtilted}) and repeat the derivation outlined above.
The only difference arises when computing the integrals over $D\uxx$ in Eq.~(\ref{eq:defLthetaintegralssc}):
\be
L_{\pm}= \int D \uxx \theta^{\pm}_{\uxx, \uz}(\{\Delta\}) \longmapsto L_{\pm}^{\gamma} =  \int D \uxx \theta^{\pm}_{\uxx, \uz}(\{\Delta\}) e^{\gamma \delta ( h_i- \lambda) }
\label{eq:defLthetaintegralssctilted}
\ee
where $h_i$ is independent of $\uxx$, and is given by Eq.~(\ref{eq:h_iasexpressofzandx}). Since we are only interested in the limit $\gamma \sim 0$, we can replace
\be
e^{\gamma \delta ( h_i- \lambda) } \sim 1 + \gamma \delta ( h_i- \lambda) = 1 + \gamma \delta ( z_1 \pm \Delta_{\hat{h}s} - \lambda) \:,
\ee
and, in the numerical iteration process, we approximate the delta function by a Gaussian function with a small variance.
Subtracting $S_{\beta}^{\gamma=0}(r)$ from $S_{\beta}^{\gamma=0.1}(r)$ and diving by $0.1$, we compute the points in the inset of Fig.~(\ref{fig:Escape}). Continuous lines are then obtained by fitting Gaussians to these points.

\end{document}